\documentclass[journal]{IEEEtran}
\usepackage{cite}

\ifCLASSINFOpdf
\else
\fi

\usepackage{xcolor}
\usepackage{graphicx}
\usepackage{amsmath, bbm}
\usepackage{amssymb}
\usepackage{amsthm}
\usepackage{booktabs}
\usepackage{multirow}
\usepackage{array,multirow}
\usepackage{cite}
\usepackage{textcomp}
\usepackage{hyperref}
\usepackage{algorithm,algorithmic}
\usepackage{cancel}
\usepackage{tikz}
\usetikzlibrary{arrows, arrows.meta}
\usepackage{setspace}

\usepackage{eso-pic}

\DeclareMathAlphabet\mathbfcal{OMS}{cmsy}{b}{n}

\begin{document}
\AddToShipoutPicture*{\small \sffamily\raisebox{27.0cm}{\hspace{2.7cm}
 Published in Front. Signal Process, Sec. Image Processing, 2022 (DOI:https://doi.org/10.3389/frsip.2022.932873)}}
 
\title{Joint Image Compression and Denoising via Latent-Space Scalability} 
\author{Saeed Ranjbar~Alvar,
        Mateen Ulhaq, Hyomin Choi
        and Ivan V. Baji\'c
        \thanks{The authors are with the School of Engineering Science, Simon Fraser University, Burnaby, BC, Canada.}}
\maketitle
\begin{abstract}

When it comes to image compression in digital cameras, denoising is traditionally performed prior to compression. However, there are applications where image noise may be necessary to demonstrate the trustworthiness of the image, such as court evidence and image forensics. This means that noise itself needs to be coded, in addition to the clean image itself. 
In this paper, we present a learning-based image compression framework where image denoising and compression are performed jointly. The latent space of the image codec is organized in a scalable manner such that the clean image can be decoded from a subset of the latent space (the base layer), while the noisy image is decoded from the full latent space at a higher rate. Using  a subset of the latent space for the denoised image allows denoising to be carried out at a lower rate. Besides providing a scalable representation of the noisy input image, performing denoising jointly with compression makes intuitive sense because noise is hard to compress; hence, compressibility is one of the criteria that may help distinguish noise from the signal.
The proposed codec is compared against established compression and denoising benchmarks, and the experiments reveal considerable bitrate savings {compared to a cascade combination of a state-of-the-art codec and a state-of-the-art denoiser}. 

 { \section{Keywords:} image denoising, image compression, deep learning, multi-task compression, scalable coding} 
\end{abstract}

\section{Introduction}
Images obtained from digital imaging sensors are degraded by the noise generated due to many factors such as lighting of the scene, sensors, shutter speed, etc. {In practice, noticeable noise is often encountered in low-light conditions, as illustrated in the Smartphone Image Denoising Dataset (SIDD)~\cite{SIDD}.} In a typical image processing pipeline, noise in the captured image is attenuated or removed before compressing the image. The noise removed in the pre-processing stage cannot be restored, and the compressed image does not carry information about the original noise. While it is a desirable feature not to have noise in the stored image for the majority of applications, the captured noise may carry useful information for certain applications, such as court evidence, image forensics, and artistic intent. For such applications, the noise needs to be preserved in the compressed image. In fact, compressed-domain denoising together with techniques to preserve the noise is part of the recent JPEG AI call for proposals~\cite{JPEG-AI_CfP}. The major drawback of encoding the noise is that it significantly increases the bitrate required for storing and transferring the images. {As an example, it is known that independent and identically distributed (iid) Gaussian source, which is a common noise model, has the worst rate-distortion performance among all the sources with the same variance~\cite{Cover_Thomas_2006}.} Another issue is that when the clean (denoised) image is needed, the denoising should be applied to the reconstructed noisy images. The additional denoising step {may increase} the run time and the complexity of the pipeline.  

To overcome the mentioned drawbacks of encoding the noisy image and performing denoising in cascade, we present a scalable multi-task image compression framework that  performs compression and denoising jointly. {We borrow the terminology from scalable video coding~\cite{scalable}, where the input video is encoded into a scalable representation consisting of a \textit{base layer} and one or more \textit{enhancement layers}, which enables reconstructing various representations of the original video - different resolutions and/or frame rates and/or qualities.} In the proposed Joint Image Compression and Denoising (JICD) framework, the encoder maps the noisy input to a latent representation that is partitioned into a base {layer} and an enhancement layer. The base layer contains the information about the clean image, while the enhancement layer contains information about noise. When the denoised image is needed, only the base layer needs to be encoded (and decoded), thereby avoiding noise coding. The enhancement layer is encoded only when the noisy input reconstruction is needed. 

The scalable design of the system provides several advantages. Since only a subset of latent features is encoded for the denoised image, the bitrate is reduced compared to using the entire latent space. Another advantage is that the noise is not completely removed from the latent features, only separated from the features corresponding to the denoised image. Therefore, when the noisy input reconstruction is needed, the enhancement features are used in addition to the base features to decode the noisy input.  The multi-task nature of the framework {means that compression and denoising are trained jointly, and it} also allows {us} to obtain both reconstructed noisy input and the corresponding denoised image in single forward pass, which reduces the complexity compared to the cascade implementation of compression and denoising. {In fact, our results demonstrate that such a system provides improved performance -- better denoising accuracy at the same bitrate -- compared to a cascade combination of a state-of-the-art codec and a state-of-the-art denoiser. 
} 

{The novel contributions of this paper are as follows:}
\begin{itemize}
  \item {We develop JICD, the first multi-task image coding framework that supports both image denoising and noisy image reconstruction. }
  \item  {JICD employs latent space scalability, such that the information about the clean image is mapped to a subset of the latent space (base layer) while noise information is mapped to the remainder (enhancement layer). }
  \item {Unlike many methods in the literature, which are either developed for a particular type of noise and/or require some noise parameter(s) in order to operate properly, the proposed JICD is capable of handling unseen noise.} 
\end{itemize}

The remainder of the paper is organized as follows. Section~\ref{sec:related} briefly describes prior work related to compression, denoising, and joint compression and denoising.  Section~\ref{sec:preliminaries} discusses the preliminaries related to learning-based multi-task image compression. {Section~\ref{sec:proposed} presents the proposed method. Section~\ref{subsec:exp} describes the experiments and analyzes the experimental results.} Finally, Section~\ref{sec:conclusion} presents concluding remarks. 

\section{Related Works}
{The proposed JICD framework is a multi-task image codec that performs image compression and denoising jointly. In this section, we briefly discuss the most relevant works related to image denoising~(Section~\ref{subsec:related_denoise}), {learning-based} image compression~(Section~\ref{subsec:related_comp}), and multi-task image compression including joint compression and denoising~(Section~\ref{subsec:related_joint}).}
\label{sec:related}
\subsection{Image Denoising}
\label{subsec:related_denoise}
State-of-the-art classical image denoising methods are based on Non-local Self Similarity (NSS). In these methods, repetitive local patterns in a noisy image are used to capture signal and noise characteristics, and perform denoising. In BM3D~\cite{BM3D}, similar patches are first found by block matching. Then, they are stacked to form a 3D block. Finally, transform-domain collaborative filtering is applied to obtain the clean patch. {\cite{BM3D-adaptive} used adaptive filtering to improve BM3D.} WNNM~\cite{WNNM} performs denoisng by applying low rank matrix approximation to the stacked noisy patches. {In~\cite{patch-NSS}, a patch group based NSS prior learning scheme to learn explicit NSS models from natural images is proposed. The denoising method in~\cite{simultaneous-NSS} used NSS priors in both the degraded images and the external clean images to perform denoising. CBM3D~\cite{CBM3D} and MCWNNM~\cite{MC-WNNM} are the extensions of BM3D and WNNM, respectively, created to handle color images.}  

More recently, (deep) learning-based denoising methods have gained popularity and surpassed the performance of classical methods. {~\cite{denoise-mlp} used a multi-layer perceptron (MLP) to achieve denoising results comparable to the state-of-the-art classic method.} Among the learning-based denoisers,   DnCNN~\cite{DnCNN} was 
the first Convolutional Neural Network (CNN) to perform blind Gaussian denoising. FFDNet~\cite{FFDNet} improved upon DnCNN by proposing a fast and flexible denoising CNN that could handle different noise levels with a single model. In~\cite{CBDNet}, noise estimation subnetwork is added prior to the CNN-based denoiser to get an accurate estimate of the noise level in the real-world noisy photographs. {A Generative Adversarial Networks (GAN)-based denoising method is proposed in~\cite{GAN-denoised}.  The mentioned works are supervised methods where  clean reference image is needed for training. In~\cite{selfsuper-denoise,self2self}, self-supervised denoising methods are proposed.} 

\subsection{{Learning-based} Image Compression}
\label{subsec:related_comp}
{In recent} years, there has been an increasing interest in the development of learning-based image codecs. {Some of the early works~\cite{toderici-variable, Minnen_ICIP2017,Johnston_CVPR2018} were based on Recurrent Neural Networks (RNNs), whose purpose was to model spatial dependence of pixels in an image. More recently, the focus has shifted to Convolutional Neural Network (CNN)-based autoencoders.} 
{ 
\cite{balle2019end} introduced Generalized
Divisive Normalization (GDN)} {as a key component of the nonlinear transform in the encoder.} {The image codec based on GDN was improved by introducing a hyperprior to capture spatial dependencies and} {take advantage} {of statistical redundancy in the entropy model~\cite{balle2018variational}. To further improve the coding gains, discretized Gaussian mixture likelihoods are used in~\cite{cheng2020image} to parameterize the distributions of latent codes.} {Most recently, this approach has been extended using advanced latent-space context modelling~\cite{Causal_context_TCSVT_2022} to achieve even better performance.}

{Most state-of-the-art learning-based image codding approaches~\cite{balle2018variational,cheng2020image,Causal_context_TCSVT_2022} train different models for different bitrates, by changing the Lagrange multiplier that trades-off rate and distortion. Such approach is meant to explore the potential of learning-based compression, rather than be used in practice as is. There has also been a considerable amount of work on variable-rate learning-based compression~\cite{toderici-variable,Conditional_AE_ICCV2019,VR_SPL2020,Sebai_MMSP2021,VR_ICASSP2022}, where a single model is able to produce multiple rate-distortion points. However, in terms of rate-distortion performance, ``fixed-rate'' approaches such as~\cite{cheng2020image,Causal_context_TCSVT_2022} currently seem to have an advantage over variable-rate ones.}

\subsection{Multi-Task Image Compression}
\label{subsec:related_joint}
The mentioned learning-based codec are single-task models, where the task is the reconstruction of the input image, just like with conventional codecs. However, the real power of learning-based codecs is their ability to be trained for multiple tasks, for example, image processing or computer vision tasks, besides the usual input reconstruction. In fact, the goal of JPEG AI standardization is to develop such a coding framework that could support multiple tasks from a common compressed representation~\cite{JPEG-AI_use_cases}. 

\cite{Choi_scalable_TIP} proposed a scalable multi-task model with multiple segments in the latent space to handle computer vision tasks in addition to input reconstruction. The concept was based on latent-space scalability~\cite{Choi_scalable_ICIP}, where the latent space is partitioned in a scalable manner, from tasks that require less information to tasks that require more information. Our JICD framework is also based on latent-space scalability~\cite{Choi_scalable_ICIP}. However, unlike {these} earlier works, the latent space is organized such that it supports image denoising from the base layer and noisy input reconstruction from the full latent space. {In other words, the tasks are different compared to these earlier works.}  

Recently,~\cite{michela} and~\cite{Satellite} developed joint image compression and denoising pipelines built upon learning-based image codecs, where the pipeline is trained to take the input noisy image, compress it, and decode a denoised image. However, with these approaches, it is not possible to reconstruct the original noisy image, hence they are not multi-task models. Our proposed JICD performs the denoising task in its base layer, but it keeps the noise information in the enhancement layer, {thereby also enabling} noisy input reconstruction {if} needed. 

\section{Prelimineries}
\label{sec:preliminaries}
In thinking about how to construct a learning-based system that can produce both the denoised image and reconstruct the noisy image, it is useful to consider the processing pipeline in which noisy image is first compressed, then decoded, and then denoising is applied to obtain the denoised image. Let $\mathbf{X}_n$ be the noisy input image. If such an image is input to a  learning-based codec~\cite{balle2019end,balle2018variational,minnen2018joint,cheng2020image}, encoding would proceed in three steps:
\begin{equation}
  \mathbfcal{Y} = g_a(\mathbf{X}_n;\phi)
\end{equation}
\begin{equation}
 \widehat{\mathbfcal{Y}} = Q(\mathbfcal{Y}) \\  
\end{equation}
\begin{equation}
  \mathbf{B} = A_E (\widehat{\mathbfcal{Y}})
\end{equation}
\noindent where $g_a$ is the  analysis transform, $\phi$ represents the parameters of $g_a$, $Q$ is the quantization function, and $\mathbf{B}$ is the bitstream obtained by applying the arithmetic encoder $A_E$ to   $\widehat{\mathbfcal{Y}}$.

The noisy input image is reconstructed at the decoder by applying the entropy decoding and synthesis transform to the encoded bitstream as:
\begin{equation}
 \widehat{\mathbfcal{Y}}  = A_D (\mathbf{B})
\end{equation}
\begin{equation}
 \widehat{\mathbf{X}}_n = g_s(\widehat{\mathbfcal{Y}};\theta)  
\end{equation}
where $A_D$ is the entropy decoder, $g_s$ and $\theta$ are the synthesis transform and its parameters, respectively. 
Then the denoised image can be obtained by applying a denoiser to the reconstructed noisy input as:
\begin{equation}
 \widehat{\mathbf{X}} = {F}(\widehat{\mathbf{X}}_n,\psi)   
\end{equation}
where $F$ and $\psi$ are the denoiser and its parameters, respectively, and $\widehat{\mathbf{X}}$ is the denoised image.

This processing pipeline forms a Markov chain $\mathbf{X}_n \to \widehat{\mathbfcal{Y}} \to \widehat{\mathbf{X}}_n \to \widehat{\mathbf{X}}$. Applying the data processing inequality (DPI)~\cite{Cover_Thomas_2006} to this Markov chain, we get
\begin{equation}
    I(\widehat{\mathbfcal{Y}};\widehat{\mathbf{X}}_n) \geq  I(\widehat{\mathbfcal{Y}};\widehat{\mathbf{X}}),
\label{eq:dpi}
\end{equation}
where $I(\cdot \, ; \, \cdot)$ is the mutual information~\cite{Cover_Thomas_2006} between two random quantities. Based on \eqref{eq:dpi} we can conclude that latent representation  $\widehat{\mathbfcal{Y}}$ carries less information about the denoised image $\widehat{\mathbf{X}}$ than it does about the noisy reconstructed image $\widehat{\mathbf{X}}_n$. Moreover, because $\widehat{\mathbf{X}}$ is obtained from $\widehat{\mathbf{X}}_n$, the information that $\widehat{\mathbfcal{Y}}$ carries about $\widehat{\mathbf{X}}$ is a subset of the information that it carries about $\widehat{\mathbf{X}}_n$.  This motivates us to structure the latent representation $\widehat{\mathbfcal{Y}}$ in such a way that only a part of it (the base layer) is used to reconstruct the denoised  image $\widehat{\mathbf{X}}$, while the whole of $\widehat{\mathbfcal{Y}}$ (base+enhancement) is used to reconstruct the noisy image $\widehat{\mathbf{X}}_n$.

\section{Proposed method}
\label{sec:proposed}
\begin{figure*}[t]
    \centering
    \centering
    \includegraphics[width=\textwidth]{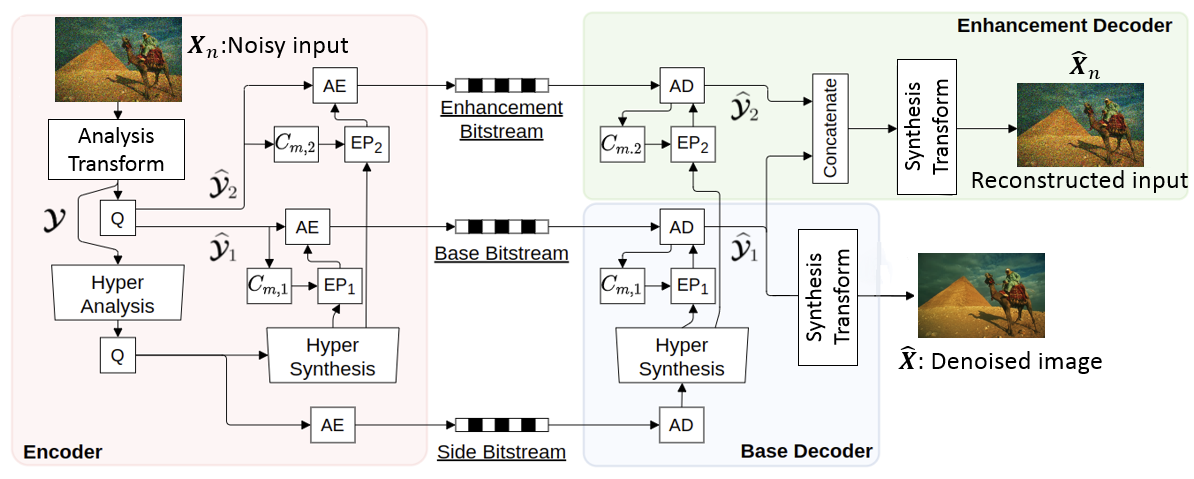}
\caption{The proposed joint compression and denoising {(JICD)} framework. AE/AD represent arithmetic encoder and decoder, respectively. $C_m$ and EP stand for the context model and entropy parameters, respectively. {Q represents the quantizer, which is simple rounding to the nearest integer. The architecture of the individual building blocks is the same as in~\cite{cheng2020image,Choi_scalable_ICIP,Choi_scalable_TIP}, but they have been retrained to support different tasks, as explained in the text.}}
\label{fig:framework}
\end{figure*}

\begin{figure*}[t]
    \centering
    \begin{minipage}[b]{0.7\linewidth}
    \centering
    \includegraphics[width=\textwidth]{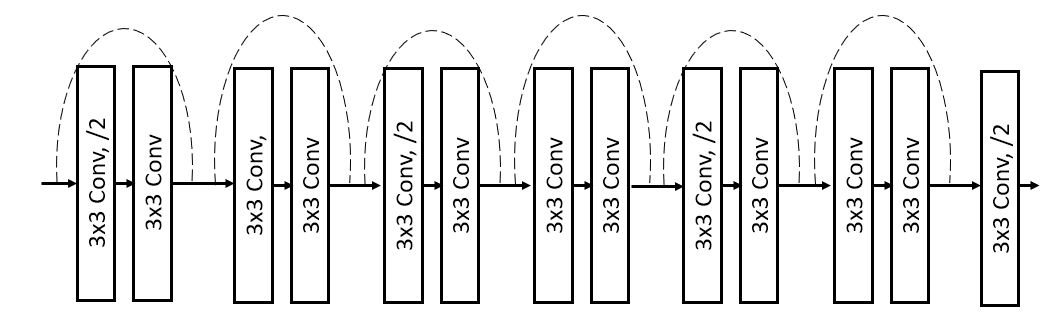}
    \centerline{(a) Analysis Transform, /2: stride=2}\medskip
    \end{minipage}
        \centering
    \begin{minipage}[b]{0.25\linewidth}
    \centering
    \vspace{50pt}
    \includegraphics[width=\textwidth]{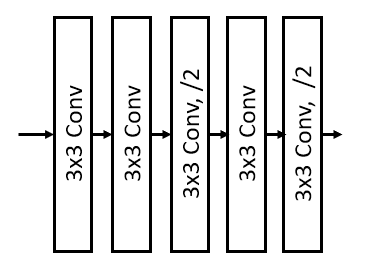}
    \centerline{(b) Hyper Synthesis}\medskip
    \end{minipage}
    \hfill
    
    \centering
    \begin{minipage}[b]{0.8\linewidth}
    \centering
    \includegraphics[width=\textwidth]{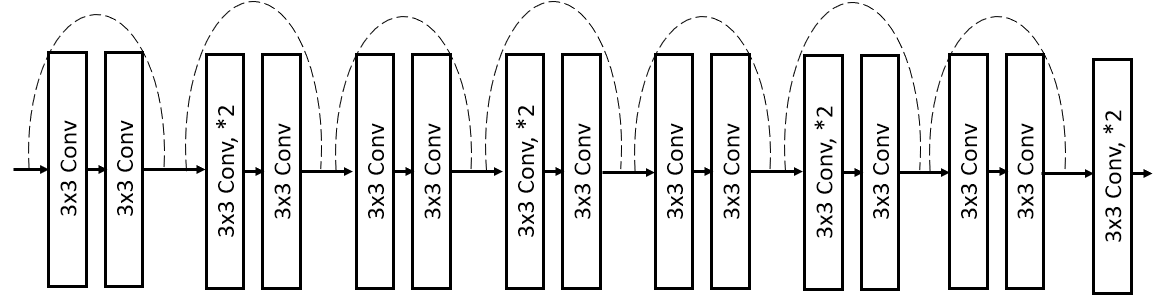}
    \centerline{(c) Synthesis Transform, *2: up-sampling by the factor of 2}\medskip
    \end{minipage}
    \hfill

    \begin{minipage}[b]{0.4\linewidth}
    \centering
    \includegraphics[width=\textwidth]{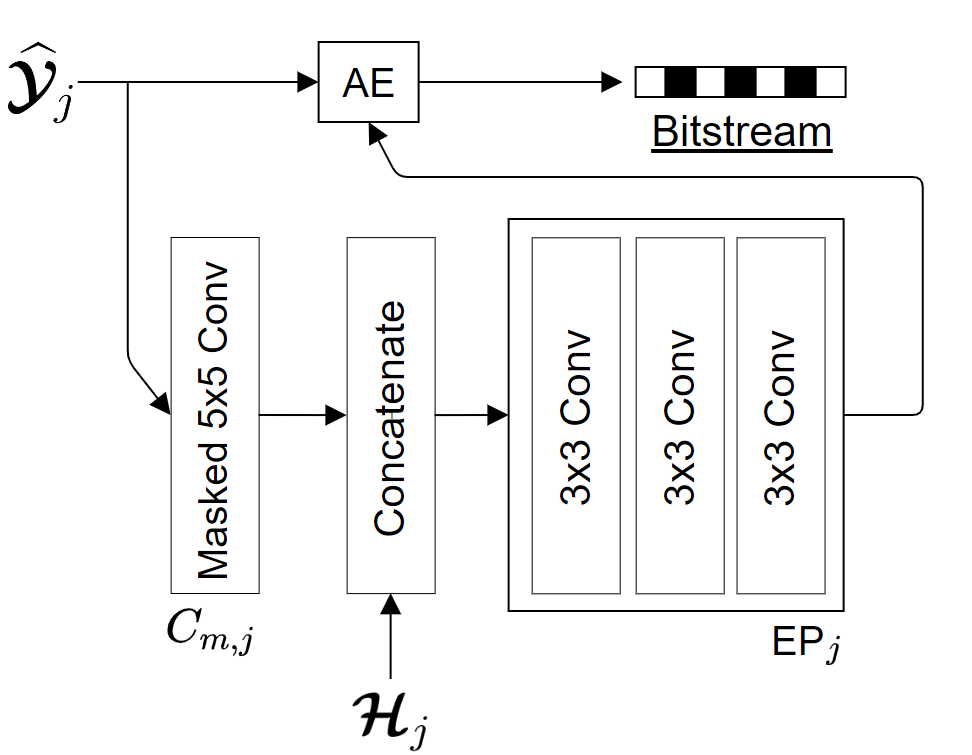}
    \centerline{(d) Context model $C_{m,j}$ and $EP_j$, $\mathbfcal{H}_j$ is the output of Hyper Synthesis }\medskip
    \end{minipage}
    \hfill
    
\caption{{The architecture of the components inside the proposed JICD framework.}}
\label{fig:arch}
\end{figure*}

The proposed joint image compression and denoising (JICD) framework consists of an encoder and two task-specific decoders, as illustrated in Fig.~\ref{fig:framework}. {The architecture of the blocks that make up the encoder and two decoders in Fig.~\ref{fig:framework} is shown in Fig.~\ref{fig:arch}. Note that the architecture of the individual building blocks (Analysis Transform, Synthesis Transform, etc.) is the same as in~\cite{cheng2020image,Choi_scalable_ICIP, Choi_scalable_TIP}, but these blocks have been retrained to support a scalable latent representation for joint compression and denoising. Specifically, compared to~\cite{cheng2020image}, our encoder is trained to produce a scalable latent space that enables both denoising and noisy input reconstruction. Compared to~\cite{Choi_scalable_ICIP,Choi_scalable_TIP}, our system is trained to support different tasks, and correspondingly the structure of the latent space and the training procedure is different.  Details of individual components are described below. 
} 

\subsection{Encoder}
\label{subsec:encoder}
The encoder {employs} an analysis transform to obtain a high fidelity latent-space representation for the input image. In addition, the encoder has blocks to efficiently encode the obtained latent-space tensor. The encoder's analysis transform is borrowed from~\cite{cheng2020image} due to its  high compression efficiency. In addition to the analysis transform, we also adopted the entropy parameter (EP) module, the context model (CTX) for arithmetic encoder/decoder (AE/AD), synthesis transform and hyper analysis/synthesis without attention layers from~\cite{cheng2020image}.

The analysis transform converts the input image $\mathbf{X}$ into $\mathbfcal{Y} \in \mathbb{R}^{N \times M \times C}$, with $C=192$ as in~\cite{minnen2018joint, cheng2020image}. {Unlike~\cite{minnen2018joint, cheng2020image},} the latent representation $\mathbfcal{Y}$ is split into two separate sub-latents $\mathbfcal{Y} = \mathbfcal{Y}_1 \cup \mathbfcal{Y}_2$, $\mathbfcal{Y}_1 \cap \mathbfcal{Y}_2=\emptyset$, where $\mathbfcal{Y}_1$ is the base layer containing $i$ channels, $\mathbfcal{Y}_1 = \{\mathbf{Y}_1, \mathbf{Y}_2,...\mathbf{Y}_i\}$, and $\mathbfcal{Y}_2$ is the enhancement layer containing $C-i$ channels, $\mathbfcal{Y}_2 = \{\mathbf{Y}_{i+1}, \mathbf{Y}_{i+2},...,\mathbf{Y}_C\}$. {This allows the latent representation to be used efficiently for multiple purposes, namely denoising (from $\mathbfcal{Y}_1$) and noisy input reconstruction (from $\mathbfcal{Y}_1 \cup \mathbfcal{Y}_2$). Since denoising requires only $\mathbfcal{Y}_1$, it can be accomplished at a lower bitrate compared to decoding the full latent space.} The sub-latents are then quantized to produce $\widehat{\mathbfcal{Y}}_1$ and $\widehat{\mathbfcal{Y}}_2$, respectively, and then coded using their respective context models to produce two independently-decodable bitstreams, as discussed in~\cite{Choi_scalable_TIP,Choi_scalable_ICIP}. The side bitstream shown in Fig.~\ref{fig:framework} is considered to be a part of the base layer and its rate is included in bitrate calculations for the base layer bitstream in the experiments.

\subsection{Decoder}
\label{subsec:decoder}
Two task-specific decoders are constructed: one for denoised image decoding and one for noisy input image reconstruction. The hyperpriors used in both decoders are reconstructed from the side bitstream which, as mentioned above, is considered to be a part of the base layer. Quantized base representation $\widehat{\mathbfcal{Y}}_1$ is reconstructed in the base decoder by decoding the base bitstream, and used to produce the denoised image $\widehat{\mathbf{X}}$. {Unlike~\cite{Choi_scalable_ICIP,Choi_scalable_TIP}, where the base layer was dedicated to object detection/segmentation, our decoder does not require latent space transformation from $\widehat{\mathbfcal{Y}}_1$ into another latent space; the synthesis transform (Fig.~\ref{fig:framework}) produces the denoised image $\widehat{\mathbf{X}}$ directly from $\widehat{\mathbfcal{Y}}_1$.} Quantized enhancement representation  $\widehat{\mathbfcal{Y}}_2$ is decoded only when noisy input reconstruction is needed. The reconstructed noisy input image $\widehat{\mathbf{X}}_n$ is produced by the second decoder using $\widehat{\mathbfcal{Y}} = \widehat{\mathbfcal{Y}}_1 \cup \widehat{\mathbfcal{Y}}_2$.

Although not pursued in this work, it is worth mentioning that the proposed JICD framework can be extended to perform various computer vision tasks as well,  such as image classification or object detection. These tasks typically require clean images, so one can think of the processing pipeline described by the following Markov chain: $\mathbf{X}_n \to \widehat{\mathbfcal{Y}}_1 \to \widehat{\mathbf{X}} \to T$, where $T$ is the output of a computer vision task, for example a class label or object bounding boxes. Applying the DPI to this Markov chain we have
\begin{equation}
    I(\widehat{\mathbfcal{Y}}_1;\widehat{\mathbf{X}}) \geq
    I(\widehat{\mathbfcal{Y}}_1;T),
\label{eq:dpi_vision}
\end{equation}
which implies that a subset of information from $\widehat{\mathbfcal{Y}}_1$ is sufficient to produce $T$. Hence, if such tasks are required, the encoder's latent space can be further partitioned by splitting $\widehat{\mathbfcal{Y}}_1$, in a manner similar to~\cite{Choi_scalable_ICIP,Choi_scalable_TIP}, to support such tasks at an even lower bitrate than our base layer.   


\subsection{Training}
\label{subsec:training}
The model is trained end-to-end with a rate-distortion Lagrangian loss function in the form of:
\begin{equation}
    \mathcal{L} = R + \lambda \cdot D,
    \label{eq:RD_loss}
\end{equation}
where $R$ is an estimate of rate rate, $D$ is the total distortion of both tasks, and $\lambda$ is the Lagrange multiplier. 
The estimated rate is affected by latent and hyper-priors as in~\cite{minnen2018joint}, 
\begin{equation}
R=\underbrace{\mathbb{E}_{x\sim p_x}\left [ -\textup{log}_{2}p_{\hat{y}}(\hat{y}) \right ]}_{\textup{latent}}+\underbrace{\mathbb{E}_{x\sim p_x}\left [ -\textup{log}_{2}p_{\hat{z}}(\hat{z}) \right ]}_{\textup{hyper-priors}},
\label{eq:rate_term}
\end{equation}
where $x$ denotes input data, $\hat{y}$ is the quantized latent data and $\hat{z}$ is the quantized hyper-prior. Total distortion $D$ is computed as the weighted average of image denoising distortion and noisy input reconstruction distortion:   
\begin{equation}
D = (1 - w) \cdot \textup{MSE}\left(\mathbf{X},\widehat{\mathbf{X}}\right) + w \cdot \textup{MSE}\left(\mathbf{X}_n,\widehat{\mathbf{X}}_n \right),
\label{eq:distortion_term}
\end{equation}
\noindent where $w$ is the trade-off factor to adjust the importance of the tasks. The automatic differentiation~\cite{auto_diff} ensures that the gradients from $D$ flow through the corresponding parameters without further modification to the back-propagation algorithm.


\section{Experimental Results}
\label{subsec:exp}
\subsection{Network Training}
The proposed multi-task model is trained from scratch using the randomly cropped $256 \times 256$ patches from the CLIC dataset~\cite{clic_dataset}. The noisy images are obtained using additive white Gaussian noise (AWGN) with three noise levels $\sigma = \{15,25,50\}$, clipping the resulting values to [0,~255] and quantizing the clipped values to mimic how noisy images are stored in practice. The batch size is set to 16. Training is ran for 300 epochs using the Adam optimizer with initial learning rate of $1\times 10^{-4}$. The learning rate is reduced by factor of $0.5$ when the training loss plateaus. We trained 6 different models by changing the value of $\lambda$ in \eqref{eq:RD_loss}. The list of different values for $\lambda$ is shown in Table~\ref{tbl:m14_lambda_values}. For all the models we used $w=0.05$ in \eqref{eq:distortion_term}. {We trained the model for the first rate point (lowest $\lambda$) from scratch. However, for the remaining rate points we fine-tune the model starting from the previous rate point's weights}

\begin{table}[htb]
\centering
\begin{tabular}{@{}ccccccc@{}}
\toprule
\\[-7pt]
Model Index & 1      & 2      & 3      & 4     & 5     & 6       \\[-7pt]     \\ 
\midrule
\\[-7pt]
$\lambda$    & 0.0035 & 0.0067 & 0.013 & 0.025 & 0.0483 & 0.09 \\[-7pt] \\
\bottomrule
\end{tabular}
\caption{$\lambda$ values used for training various models. Higher $\lambda$ leads to higher qualities and higher bitrates. }
\label{tbl:m14_lambda_values}
\end{table}

We trained models under two different settings. In the first setting, a given model is trained for each noise level. For this case, the number of enhancement channels $C-i$ is chosen according to the strength of the noise. For stronger noise, we allocate more channels to the enhancement layer, so that it can capture enough information to reconstruct the noise. 
The number of enhancement channels is reduced as the noise gets weaker. Specifically, the number of enhancement channels is empirically set to 32, 12, and 2 for $\sigma=50$, $\sigma=25$, and $\sigma=15$, respectively. 
The second training setting is to train a single model with different noise levels $\sigma \in \{50,25,15\}$ simultaneously, and use the final trained model to perform denoising for all noise levels. This is beneficial when the noise level information is not given. In this model, we used 180 base channels and 12 enhancement channels. $\sigma$ at each training iteration is uniformly chosen from $\{50,25,15\}$.     

\subsection{Data}
To evaluate the performance of the proposed JICD framework, four color image datasets are used: 1) CBSD68~\cite{CBSD}, 2) Kodak24~\cite{kodak}, 3) McMaster~\cite{McMaster} and 4) JPEG AI testset~\cite{JPEG-AI_use_cases}, which is used in the JPEG AI exploration experiments.  The mentioned datasets contain 68, 24, 18, and 16 images, respectively. The resolution of the images in the Kodak24 and McMaster dataset is fixed to $500\times500$. CBS68 dataset contains the lowest-resolution images among the four datasets, with the height and width of images ranging between 321 and 481. The images in the JPEG AI testset are high-resolution images with the height varying between 872 and 2456 pixels and width varying between 1336 and 3680 pixels. The results are reported for two sets of noisy images. In the first set, we added synthesized AWGN to the testing images with three noise levels: $\sigma = \{15,25,50\}$ and tested the results with the quantized noisy images. In the second set, we used the synthesized noise obtained from the noise simulator in~\cite{PNG}, which was also used to generate the final test images for the denoising tasks in the ongoing JPEG AI standardization. This type  of noise was not used during the training of the proposed JICD framework. Hence, the goal of testing with this second set of images is to evaluate how well the proposed JICD generalizes to the noise that is not seen during the training.

\subsection{Baselines}
\label{subsec:exp_baselines}
The denoising performance of the proposed JICD framework is compared against well-established baselines: CBM3D~\cite{CBM3D} and FFDNet~\cite{FFDNet}. CBM3D is a NSS-based denoising method, and FFDNet belongs to the learning-based denoising category.  FFDNet was trained using AWGN with different noise levels during the training. At inference time, FFDNet needs the variance of the noise as input. FFDNet-clip~\cite{FFDNet} is a version of FFDNet that is trained with quantized noisy images. Since our focus is on practical settings with quantized noisy images, we used FFDNet-clip as a baseline in the experiments. We also tested the DRUNet denoiser~\cite{drunet}, which is one of the latest state-of-the-art denoisers. DRUNet assumes that the noise is not quantized, and when tested with quantized noise, it performs worse than FFDNet-clip. As a result, we did not include it in the experiments.

Two baselines are established by applying CBM3D and FFDNet-clip directly on noisy images, without compression. However, to assess the interaction of compression and denoising, we establish one more baseline. In this third baseline, the noisy image is first compressed using the end-to-end image compression model from~\cite{cheng2020image} (the ``Cheng model'') with an implementation from CompressAI~\cite{begaint2020compressai}, and then decoded. Then FFDNet-clip is used to denoise the decoded noisy image. We call this cascade denoising approach as Cheng+FFDNet-clip. It is worth mentioning that Cheng+FFDNet-clip, similar to the proposed JICD framework, is able to obtain both the reconstructed noisy images and denoised images, hence it could be considered a multi-task approach.

\subsection{Experiments on AWGN Removal}
\label{subsec:exp_AWG}
\begin{figure*}[t]
    \centering
    \begin{minipage}[b]{0.45\linewidth}
    \centering
    \includegraphics[width=\textwidth]{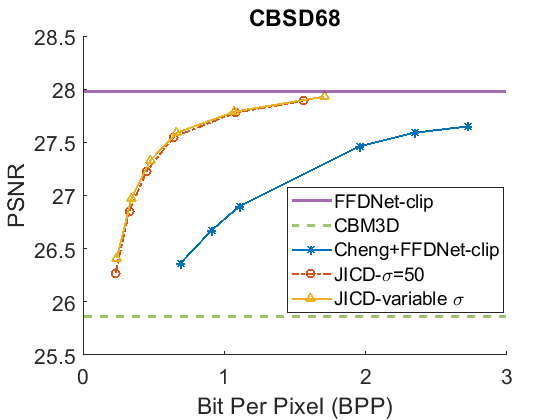}
    \centerline{(a)}\medskip
    \end{minipage}
    \begin{minipage}[b]{0.45\linewidth}
    \centering
    \includegraphics[width=\textwidth]{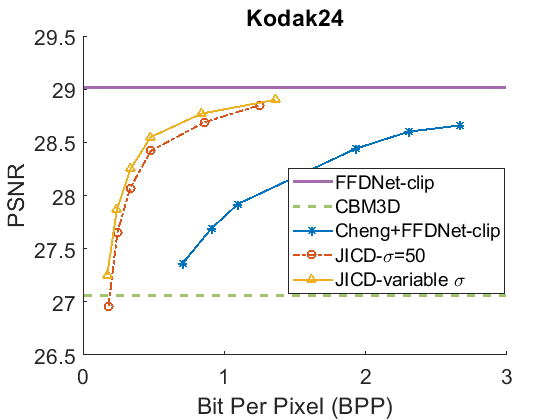}
    \centerline{(b)}\medskip
    \end{minipage}
    \hfill
    
    \centering
    \begin{minipage}[b]{0.45\linewidth}
    \centering
    \includegraphics[width=\textwidth]{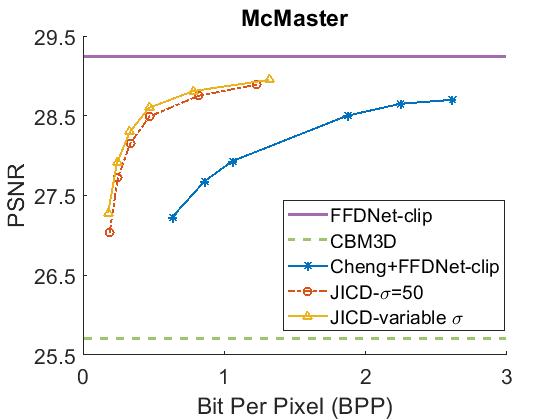}
    \centerline{(c)}\medskip
    \end{minipage}
    \begin{minipage}[b]{0.45\linewidth}
    \centering
    \includegraphics[width=\textwidth]{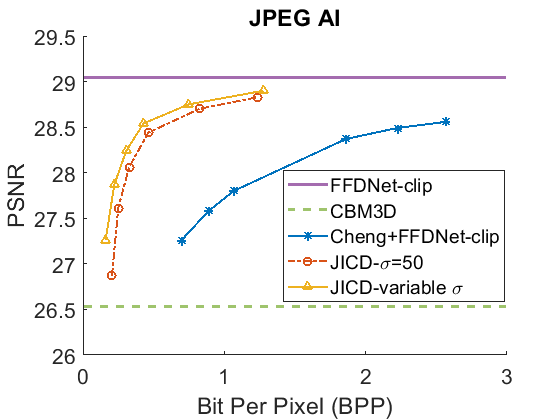}
    \centerline{(d)}\medskip
    \end{minipage}
    \hfill
\caption{Denoising rate-PSNR curves for $\sigma=50$: (a)~CBSD68, (b)~Kodak24, (c)~McMaster, (d)~JPEG AI  }
\label{fig:results_sigma_50}
\end{figure*}

\begin{figure*}[t]
    \centering
    \begin{minipage}[b]{0.45\linewidth}
    \centering
    \includegraphics[width=\textwidth]{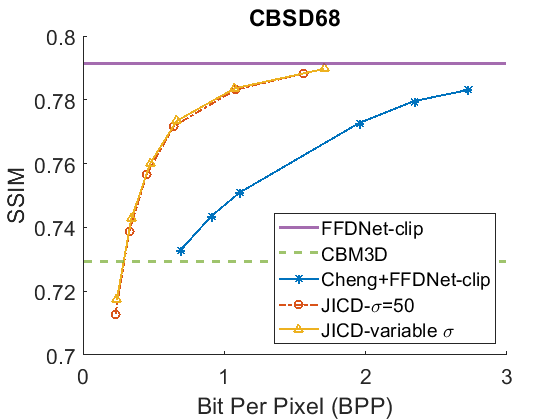}
    \centerline{(a)}\medskip
    \end{minipage}
    \begin{minipage}[b]{0.45\linewidth}
    \centering
    \includegraphics[width=\textwidth]{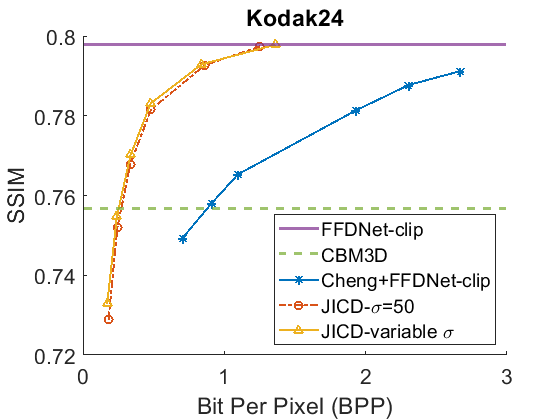}
    \centerline{(b)}\medskip
    \end{minipage}
    \hfill
    
    \centering
    \begin{minipage}[b]{0.45\linewidth}
    \centering
    \includegraphics[width=\textwidth]{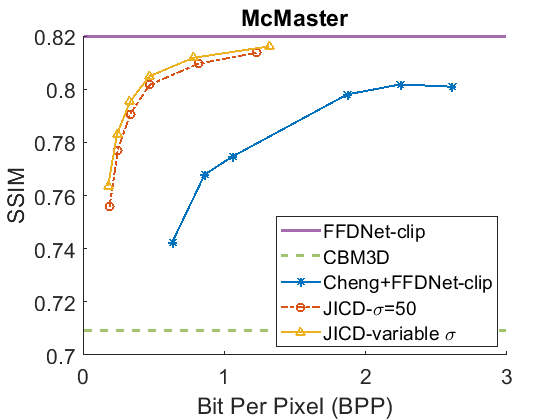}
    \centerline{(c)}\medskip
    \end{minipage}
    \begin{minipage}[b]{0.45\linewidth}
    \centering
    \includegraphics[width=\textwidth]{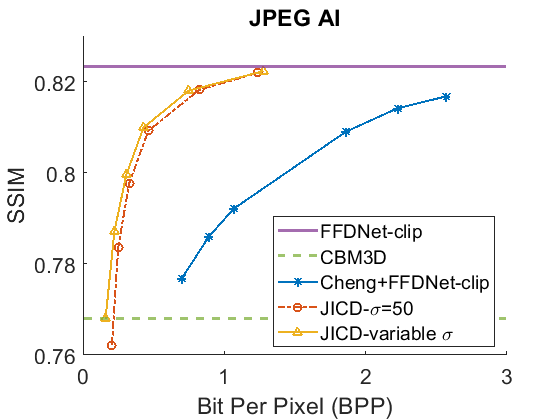}
    \centerline{(d)}\medskip
    \end{minipage}
    \hfill
\caption{{Denoising rate-SSIM curves for $\sigma=50$: (a)~CBSD68, (b)~Kodak24, (c)~McMaster, (d)~JPEG AI}}
\label{fig:results_sigma_50_SSIM}
\end{figure*}

We evaluate the baselines and the proposed JICD method using the quantized noisy images obtained using AWGN with three noise levels, $\sigma \in \{15,25,50\}$. The test results with the strongest noise ($\sigma=50$) across the four datasets (CBSD68, Kodak24, McMaster, and JPEG AI) are shown Fig.~\ref{fig:results_sigma_50} {in terms of rate vs. Peak Signal-to-Noise Ratio (PSNR) and in Fig.~\ref{fig:results_sigma_50_SSIM} in terms of rate vs. Structural Similarity Index Measure (SSIM)}. The horizontal lines in the figure correspond to applying CBM3D and FFDNet-clip to the raw (uncompressed) noisy images. The blue curve shows the results for Cheng+FFDNet-clip. The six points on this curve correspond to the six Cheng models from CompressAI~\cite{begaint2020compressai}.  For JICD, two curves are shown. The orange curve shows the results obtained from the models trained for $\sigma=50$ with 160 base feature channels and 32 enhancement channels. The yellow curve corresponds to the results obtained using the model that was trained with variable $\sigma$ values 
and has 180 base and 12 enhancement channels. The six points on the orange and yellow curves correspond to the six JICD models we trained with $\lambda$ values shown in Table~\ref{tbl:m14_lambda_values}. 


As seen in Fig.~\ref{fig:results_sigma_50}, for $\sigma=50$, the quality of the images denoised by CBM3D is considerably lower compared to those obtained using FFDNet-clip. It was shown in~\cite{FFDNet} that CBM3D and FFDNet-clip achieve comparable performance for non-quantized noisy images. Our results show that CBM3D's performance is degraded when the noise deviates (due to clipping and quantization) from the assumed model, at least at high noise levels. 

The comparison of the results obtained by JICD and Cheng+FFDNet-clip reveal that JICD is able to reduce the bitrate substantially while achieving the same denoising performance as Cheng+FFDNet-clip. This is due to the fact that the Cheng model allocates the entire latent representation to noisy input reconstruction, whereas the proposed method uses a subset of the latent features to perform denoising. The results of JICD trained with variable $\sigma$ are also shown in the curves. Since the number of base channels is larger in this model compared to the model trained for $\sigma=50$, its denoising performance is improved. 

To summarize the differences between the performance-rate curves, we compute Bj{\o}ntegaard Delta-rate (BD-rate)~\cite{bd_br}. The BD-rate of the proposed JICD compared to  Cheng+FFDNet-clip on the four datasets is given in the first two rows of Table~\ref{tbl:bd_rate} {for PSNR, and Table~\ref{tbl:bd_rate_SSIM} for SSIM}. It can be seen that the proposed method achieves up to 80.2\% BD-rate savings compared to Cheng+FFDNet-clip.  Both JICD and Cheng+FFDNet-clip denoising methods outperform CBMD3D for all the tested rate points at $\sigma=50$. Using the proposed JICD method, we are able to denoise images at a quality close to what FFDNet-clip achieves on raw images, and at the same time compress the input.

\renewcommand{\arraystretch}{1}
\begin{table*}[t]
\centering
\begin{tabular}{|c|crrrr|}
\hline
\multicolumn{1}{|c|}{Noise type}                                              &   Model                 & \multicolumn{1}{c}{CBSD68} & \multicolumn{1}{c}{Kodak24} & \multicolumn{1}{l}{McMaster} & JPEG AI \\ \hline \hline
\multirow{2}{*}{AWGN $\sigma=50$}                                        & $\sigma=50$        &   $-69.28$\%    &  $-72.91$\%      &    $-72.69$\%         &  $-74.45$\%       \\
                                                                    & variable  $\sigma$ &   $-70.66$\%    &  $-77.27$\%      &    $-76.55$\%         &  $-80.20$\%       \\ \hline
\multirow{2}{*}{AWGN $\sigma=25$}                                        & $\sigma=25$        &   $-30.58$\%    &  $-41.00$\%      &    $-33.18$\%         &   $-45.13$\%      \\
                                                                    & variable  $\sigma$ &   $-30.28$\%    &  $-42.61$\%      &    $-33.52$\%         &   $-45.77$\%      \\ \hline
\multirow{2}{*}{AWGN $\sigma=15$}                                        & $\sigma=15$        &   $1.07$\%      &   $-11.99$\%     &    $-4.95$\%          &   $-15.82$\%      \\
                                                                    & variable  $\sigma$ &   $8.00$\%      &   $-2.99$\%      &     $9.22$\%          &    $-5.78$\%     \\ \hline
\begin{tabular}[c]{@{}c@{}}Practical noise\\ simulator\end{tabular} & variable  $\sigma$ &  $-23.25$\%     &   $-33.83$\%     &   $-21.51$\%          &   $-23.42$\%
      \\ \hline
\end{tabular}
\caption{The PSNR-based BD-rate of the proposed JICD compared to Cheng+FFDNet-clip on the image denoising task.}
\label{tbl:bd_rate}
\end{table*}
\renewcommand{\arraystretch}{1}

\renewcommand{\arraystretch}{1.2}
\begin{table*}[t]
\centering
\begin{tabular}{|c|crrrr|}
\hline
\multicolumn{1}{|c|}{Noise type}                                              &   Model                 & \multicolumn{1}{c}{CBSD68} & \multicolumn{1}{c}{Kodak24} & \multicolumn{1}{l}{McMaster} & JPEG AI \\ \hline \hline
\multirow{2}{*}{AWGN $\sigma=50$}                                        &      $\sigma=50$        &   $-63.34$\%    &  $-72.08$\%      &  $-72.29$\%          &  $-72.89$\%       \\
                                                                    & variable  $\sigma$ &   $-64.09$\%    &  $-73.64$\%      &    $-75.57$\%         &  $-76.43$\%       \\ \hline
\multirow{2}{*}{AWGN $\sigma=25$}                                        & $\sigma=25$        &   $-24.14$\%    &  $-38.99$\%      &    $-53.40$\%         &   $-43.11$\%      \\
                                                                    & variable  $\sigma$ &   $-24.45$\%    &  $-39.86$\%      &    $-52.21$\%         &   $-43.97$\%      \\ \hline
\multirow{2}{*}{AWGN $\sigma=15$}                                        & $\sigma=15$        &   $4.52$\%      &   $-11.85$\%     &    $-21.57$\%          &   $-15.32$\%      \\
                                                                    & variable  $\sigma$ &   $9.14$\%      &   $-5.96$\%      &     $-8.60$\%          &    $-8.78$\%     \\ \hline
\begin{tabular}[c]{@{}c@{}}Practical noise\\ simulator\end{tabular} & variable  $\sigma$ &  $-15.83$\%     &   $-27.66$\%     &   $-28.18$\%          &   $-37.72$\%
      \\ \hline
\end{tabular}
\caption{{The SSIM-based BD-rate of the proposed JICD compared to Cheng+FFDNet-clip on the image denoising task.} }
\label{tbl:bd_rate_SSIM}
\end{table*}
\renewcommand{\arraystretch}{1}

\begin{figure*}[tb!]
    \centering
    \begin{minipage}[b]{0.45\linewidth}
    \centering
    \includegraphics[width=\textwidth]{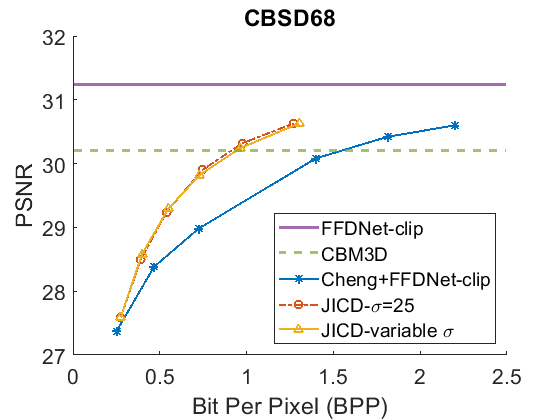}
    \centerline{(a)}\medskip
    \end{minipage}
    \begin{minipage}[b]{0.45\linewidth}
    \centering
    \includegraphics[width=\textwidth]{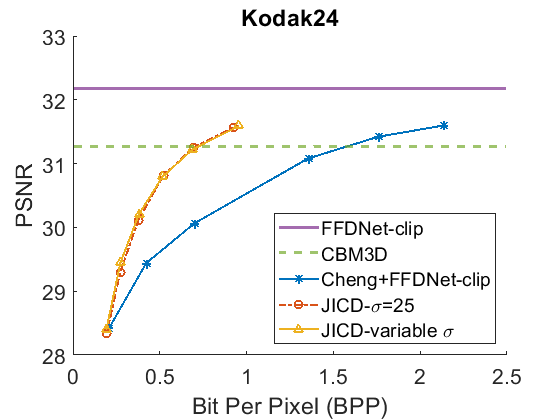}
    \centerline{(b)}\medskip
    \end{minipage}
    \hfill
    
    \centering
    \begin{minipage}[b]{0.45\linewidth}
    \centering
    \includegraphics[width=\textwidth]{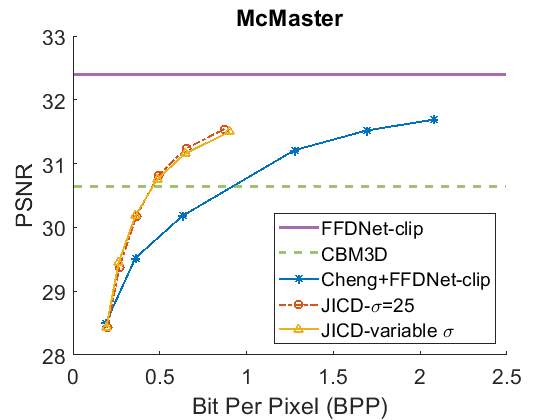}
    \centerline{(c)}\medskip
    \end{minipage}
    \begin{minipage}[b]{0.45\linewidth}
    \centering
    \includegraphics[width=\textwidth]{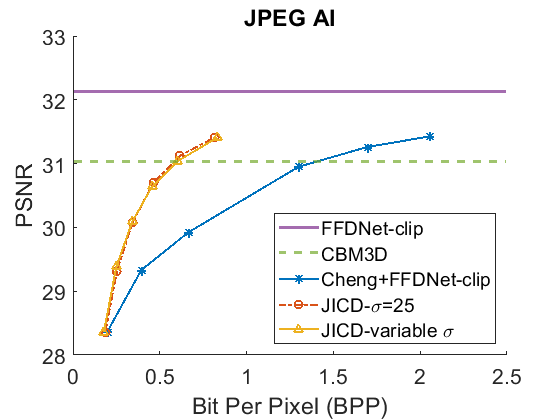}
    \centerline{(d)}\medskip
    \end{minipage}
    \hfill
\caption{Denoising rate-PSNR curves for $\sigma=25$: (a)~CBSD68, (b)~Kodak24, (c)~McMaster, (d)~JPEG AI}
\label{fig:results_sigma_25}
\end{figure*}

\begin{figure*}[tb!]
    \centering
    \begin{minipage}[b]{0.45\linewidth}
    \centering
    \includegraphics[width=\textwidth]{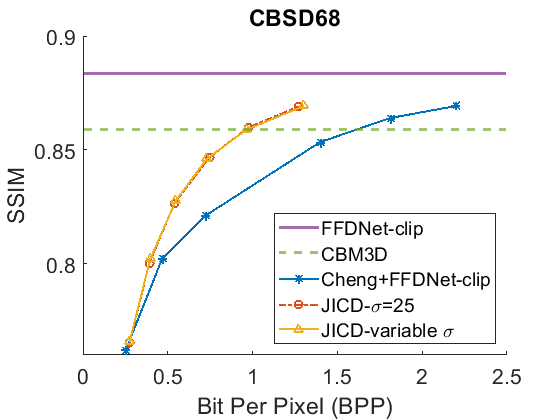}
    \centerline{(a)}\medskip
    \end{minipage}
    \begin{minipage}[b]{0.45\linewidth}
    \centering
    \includegraphics[width=\textwidth]{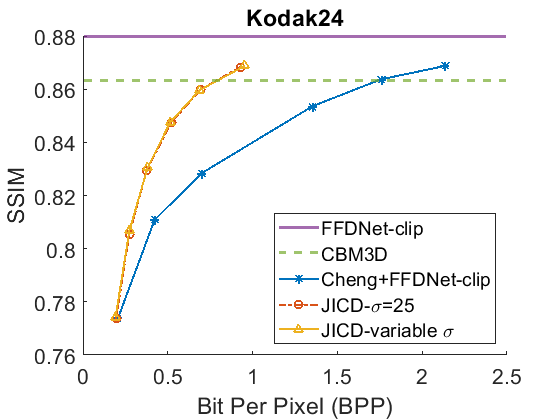}
    \centerline{(b)}\medskip
    \end{minipage}
    \hfill
    
    \centering
    \begin{minipage}[b]{0.45\linewidth}
    \centering
    \includegraphics[width=\textwidth]{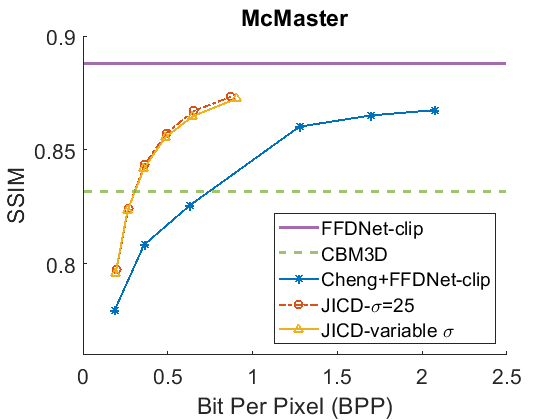}
    \centerline{(c)}\medskip
    \end{minipage}
    \begin{minipage}[b]{0.45\linewidth}
    \centering
    \includegraphics[width=\textwidth]{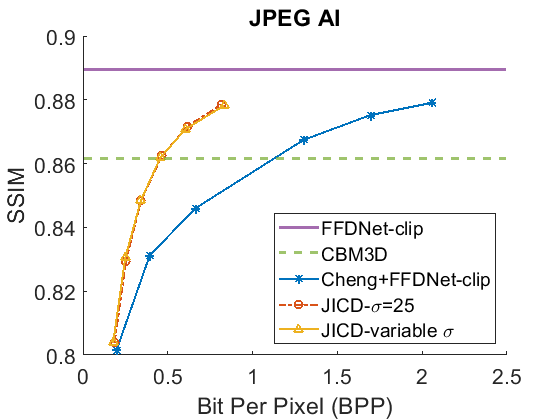}
    \centerline{(d)}\medskip
    \end{minipage}
    \hfill
\caption{{Denoising rate-SSIM curves for $\sigma=25$: (a)~CBSD68, (b)~Kodak24, (c)~McMaster, (d)~JPEG AI}}
\label{fig:results_sigma_25_SSIM}
\end{figure*}

We repeat the denoising experiment for $\sigma=25$, and the results are shown in Fig.~\ref{fig:results_sigma_25} {for PSNR and Fig.~\ref{fig:results_sigma_25_SSIM} for SSIM}. As seen in the figures, the gap between the CBM3D and FFDNet-clip performance is now reduced, and the compression-based methods now outperform CBM3D only at the higher rates. The gap between the curves corresponding to JICD and Cheng+FFDNet-clip is also reduced. However, JICD still achieves a considerable BD-rate saving 
compared to Cheng+FFDNet-clip, as shown in the third row of Table~\ref{tbl:bd_rate} and {Table~\ref{tbl:bd_rate_SSIM}}. JICD trained with variable $\sigma$ has slightly better PSNR performance compared to the noise-specific model on three datasets, and a slightly worse performance (by 0.3\%) on the low-resolution CBSD68 dataset. 


\begin{figure*}[tb!]
    \centering
    \begin{minipage}[b]{0.45\linewidth}
    \centering
    \includegraphics[width=\textwidth]{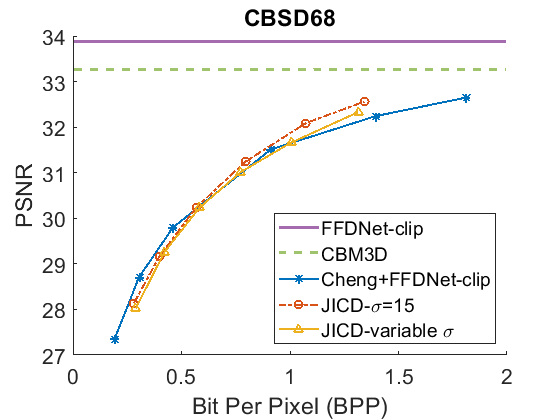}
    \centerline{(a)}\medskip
    \end{minipage}
    \begin{minipage}[b]{0.45\linewidth}
    \centering
    \includegraphics[width=\textwidth]{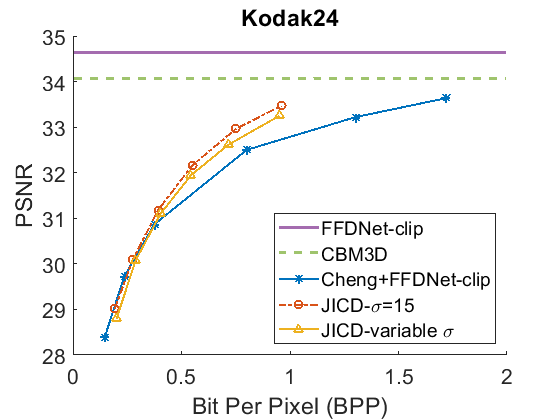}
    \centerline{(b)}\medskip
    \end{minipage}
    \hfill
    
    \centering
    \begin{minipage}[b]{0.45\linewidth}
    \centering
    \includegraphics[width=\textwidth]{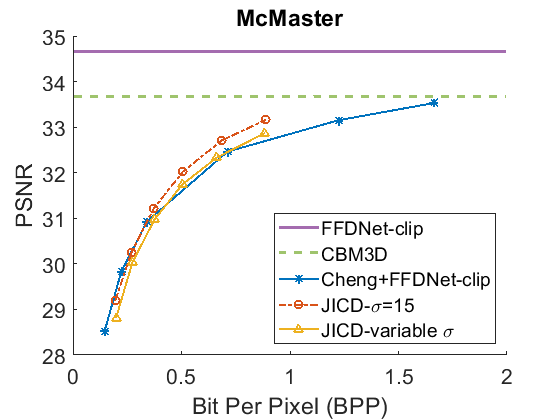}
    \centerline{(c)}\medskip
    \end{minipage}
    \begin{minipage}[b]{0.45\linewidth}
    \centering
    \includegraphics[width=\textwidth]{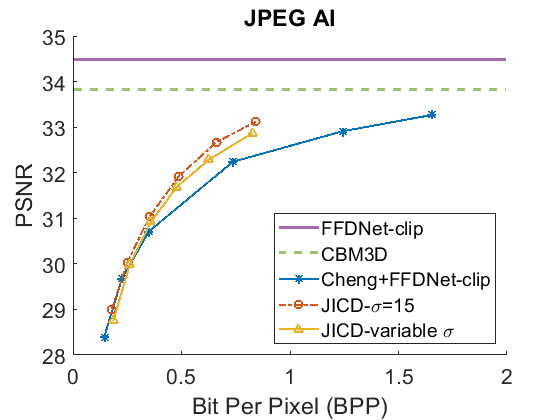}
    \centerline{(d)}\medskip
    \end{minipage}
    \hfill
\caption{Denoising rate-PSNR curves for $\sigma=15$: (a)~CBSD68, (b)~Kodak24, (c)~McMaster, (d)~JPEG AI}
\label{fig:results_sigma_15}
\end{figure*}

\begin{figure*}[tb!]
    \centering
    \begin{minipage}[b]{0.45\linewidth}
    \centering
    \includegraphics[width=\textwidth]{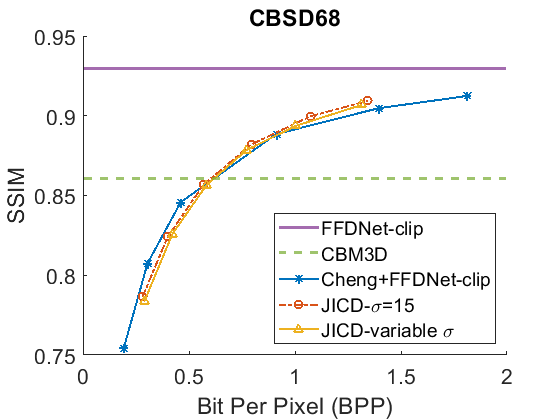}
    \centerline{(a)}\medskip
    \end{minipage}
    \begin{minipage}[b]{0.45\linewidth}
    \centering
    \includegraphics[width=\textwidth]{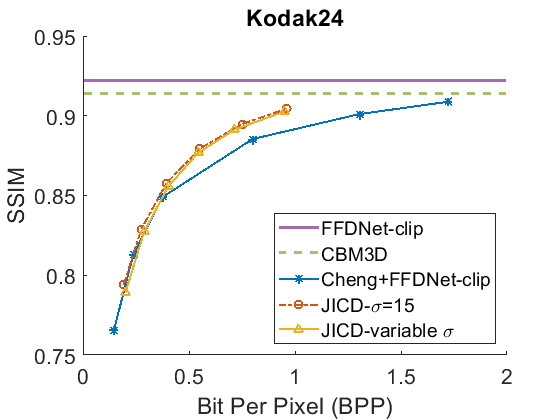}
    \centerline{(b)}\medskip
    \end{minipage}
    \hfill
    
    \centering
    \begin{minipage}[b]{0.45\linewidth}
    \centering
    \includegraphics[width=\textwidth]{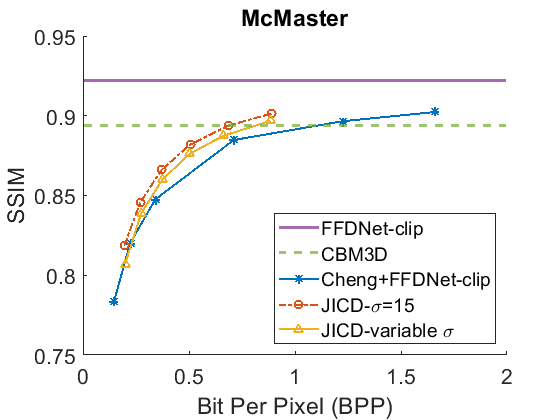}
    \centerline{(c)}\medskip
    \end{minipage}
    \begin{minipage}[b]{0.45\linewidth}
    \centering
    \includegraphics[width=\textwidth]{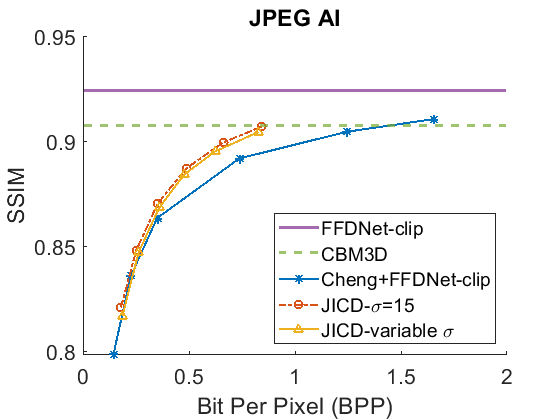}
    \centerline{(d)}\medskip
    \end{minipage}
    \hfill
\caption{{Denoising rate-SSIM curves for $\sigma=15$: (a)~CBSD68, (b)~Kodak24, (c)~McMaster, (d)~JPEG AI}}
\label{fig:results_sigma_15_SSIM}
\end{figure*}

At the lowest noise level ($\sigma=15$), the gap between CBM3D and FFDNet-clip shrinks further. It can be seen in the denoised rate-PSNR curves in Fig.~\ref{fig:results_sigma_15} {and rate-SSIM curves in Fig.~\ref{fig:results_sigma_15_SSIM}} that when the noise is weak, applying denoising to the raw images achieves high PSNR, and the compression-based methods cannot outperform either CBM3D, or FFDNet-clip at the tested rates. The gap between JICD and Cheng+FFDNet-clip curves is also reduced compared to the higher noise levels. This can also be be seen from the BD-rates in the fourth row of Table~\ref{tbl:bd_rate} and {Table~\ref{tbl:bd_rate_SSIM}}.  JICD trained for $\sigma=15$ outperforms Cheng+FFDNet-clip on three datasets, but it suffers a 1\% ({4.5\% for SSIM}) loss  on the low-resolution CBSD68. 

As seen above, the performance of the proposed JICD framework is lower on the low-resolution CBSD68 dataset than on other datasets. The reason is the following. The processing pipeline USED in JICD expects the input dimensions to be multiples of 64. For images whose dimensions do not satisfy this requirement, the input is padded up to the nearest multiple of 64. At low resolutions, the padded area may be somewhat large in relation to the original image, which causes noticeable performance degradation. At high resolutions, the padded area is insignificant compared to the original image, and the impact on JICD's performance is correspondingly smaller. It is worth mentioning that for $\sigma=15$, the JICD trained with variable $\sigma$ has a weaker denoising performance compared to the model trained specifically for $\sigma=15$. This is because the number of base channels in the variable-$\sigma$ model (180) is smaller than the number of base channels in the noise-specific model (190). At low noise levels, fewer channels are needed to hold noise information, which means the number of base channels could be higher. Hence,  the structure chosen for the noise-specific model is better suited for this case. However, we show in the next subsection that the variable-$\sigma$ model is more useful when the noise parameters are not known. 


\subsection{Experiments on Unseen Noise Removal}
\label{subsec:exp_unseen}
\subsubsection{Image Compression and Denoising}
\label{subsubsec:exp_unseen_denoise}
The proposed JICD denoiser and the baselines are also tested with the noise that was not used in the training. The purpose of this experiment is to evaluate how well the denoisers are able to handle unseen noise. To generate unseen noise, we used the noise simulator from~\cite{PNG}. This noise simulator, which we subsequently refer to as `practical noise simulator,' was created by fitting the Poissonian-Gaussian noise model~\cite{Foi} to the noise from the Smartphone Image Denoising Dataset (SIDD)~\cite{SIDD}. It is worth mentioning that this noise simulator is used in the evaluation of the image denoising task in JPEG AI standardization. 

For this experiment we use the JICD  model trained with variable $\sigma$. One advantage of this model is that, unlike some of the baselines, it does not require any additional input or noise information, besides the noisy image. On the other hand, FFDNet needs $\sigma$ to perform denoising. In the experiment, the $\sigma$ is estimated for each image by computing the standard deviation of the difference between the noisy test image and the corresponding clean image. 


\begin{figure*}[tb!]
    \centering
    \begin{minipage}[b]{0.45\linewidth}
    \centering
    \includegraphics[width=\textwidth]{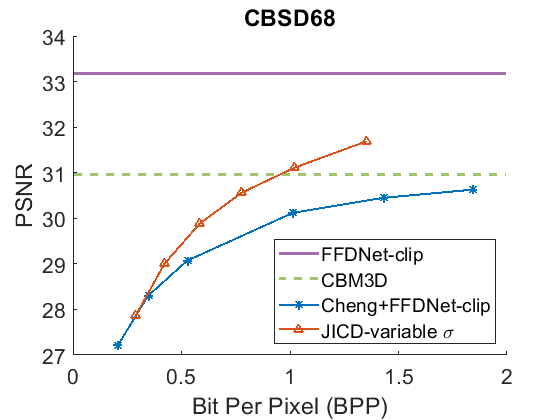}
    \centerline{(a)}\medskip
    \end{minipage}
    \begin{minipage}[b]{0.45\linewidth}
    \centering
    \includegraphics[width=\textwidth]{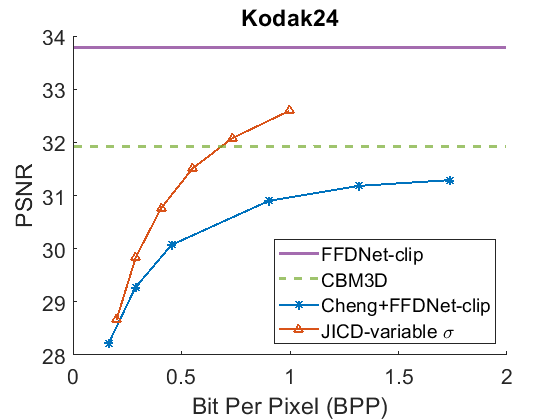}
    \centerline{(b)}\medskip
    \end{minipage}
    \hfill
    
    \centering
    \begin{minipage}[b]{0.45\linewidth}
    \centering
    \includegraphics[width=\textwidth]{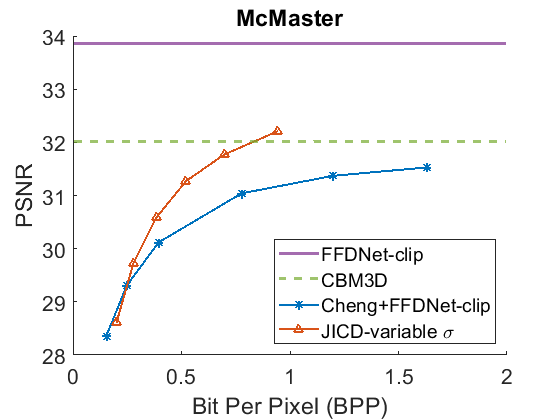}
    \centerline{(c)}\medskip
    \end{minipage}
    \begin{minipage}[b]{0.45\linewidth}
    \centering
    \includegraphics[width=\textwidth]{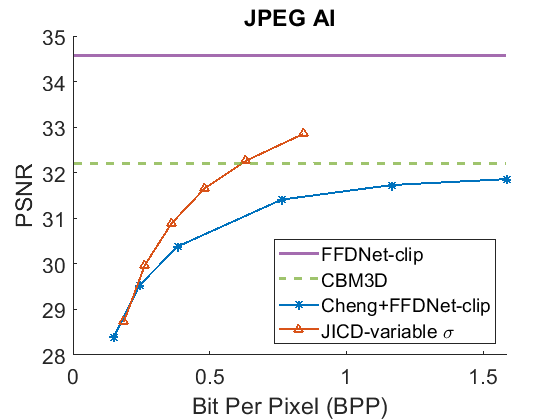}
    \centerline{(d)}\medskip
    \end{minipage}
    \hfill
\caption{Denoising rate-PSNR curves for the unseen noise: (a)~CBSD68, (b)~Kodak24, (c)~McMaster, (d)~JPEG AI}
\label{fig:results_sigma_png}
\end{figure*}

\begin{figure*}[tb!]
    \centering
    \begin{minipage}[b]{0.45\linewidth}
    \centering
    \includegraphics[width=\textwidth]{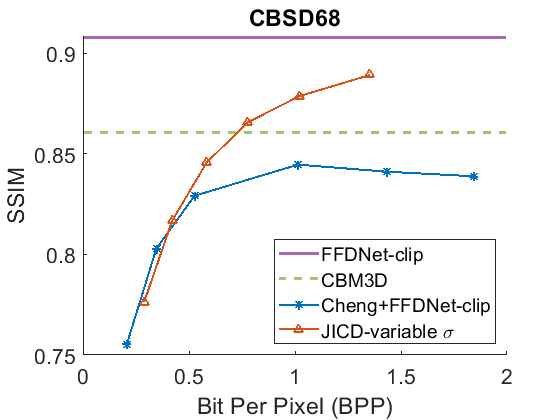}
    \centerline{(a)}\medskip
    \end{minipage}
    \begin{minipage}[b]{0.45\linewidth}
    \centering
    \includegraphics[width=\textwidth]{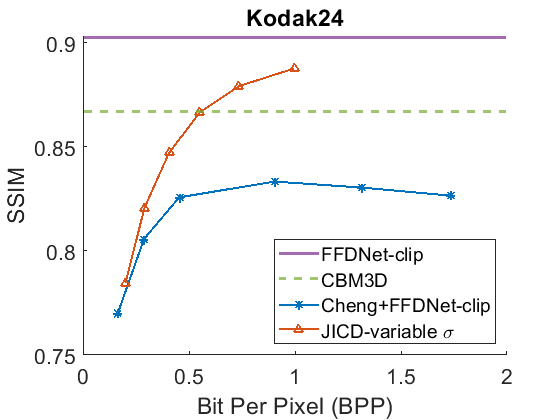}
    \centerline{(b)}\medskip
    \end{minipage}
    \hfill
    
    \centering
    \begin{minipage}[b]{0.45\linewidth}
    \centering
    \includegraphics[width=\textwidth]{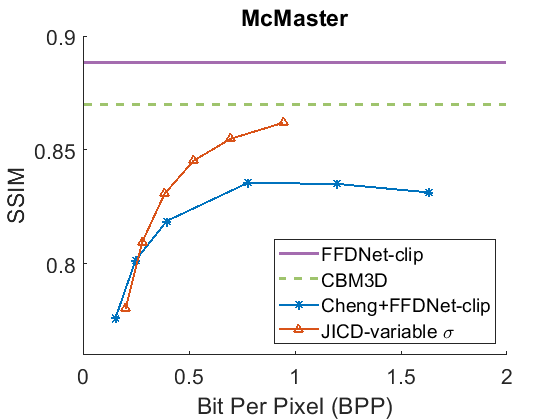}
    \centerline{(c)}\medskip
    \end{minipage}
    \begin{minipage}[b]{0.45\linewidth}
    \centering
    \includegraphics[width=\textwidth]{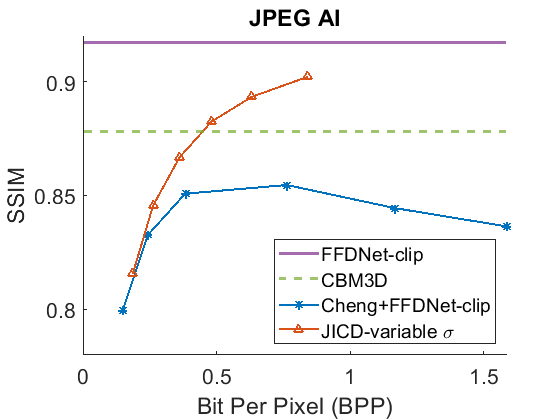}
    \centerline{(d)}\medskip
    \end{minipage}
    \hfill
\caption{{Denoising rate-SSIM curves for the unseen noise: (a)~CBSD68, (b)~Kodak24, (c)~McMaster, (d)~JPEG AI}}
\label{fig:results_sigma_png_SSIM}
\end{figure*}

The denoising rate-PSNR {and rate-SSIM} curves are illustrated in Fig.~\ref{fig:results_sigma_png} {and Fig.~\ref{fig:results_sigma_png_SSIM}, respectively}. Since the variance of the noise obtained from the practical noise simulator is not large, the PSNR range of the denoised images is close to that observed in the AWGN experiments with  $\sigma=15$ and $\sigma=25$. 
The results indicate that JICD achieves better denoising performance compared to Cheng+FFDNet-clip across  all four datasets. Moreover, at higher bitrates (1 bpp and above), JICD outperforms CBM3D applied to uncompressed noisy images. BD-rate results are summarized in the last row of Table~\ref{tbl:bd_rate} and {Table~\ref{tbl:bd_rate_SSIM}}. It is seen in the table that JICD achieves $15$-$30\%$ gain over Cheng+FFDNer-clip across the four datasets.

\begin{figure*}[t!]
    \centering
    \begin{minipage}[b]{0.45\linewidth}
    \centering
    \includegraphics[width=0.95\textwidth]{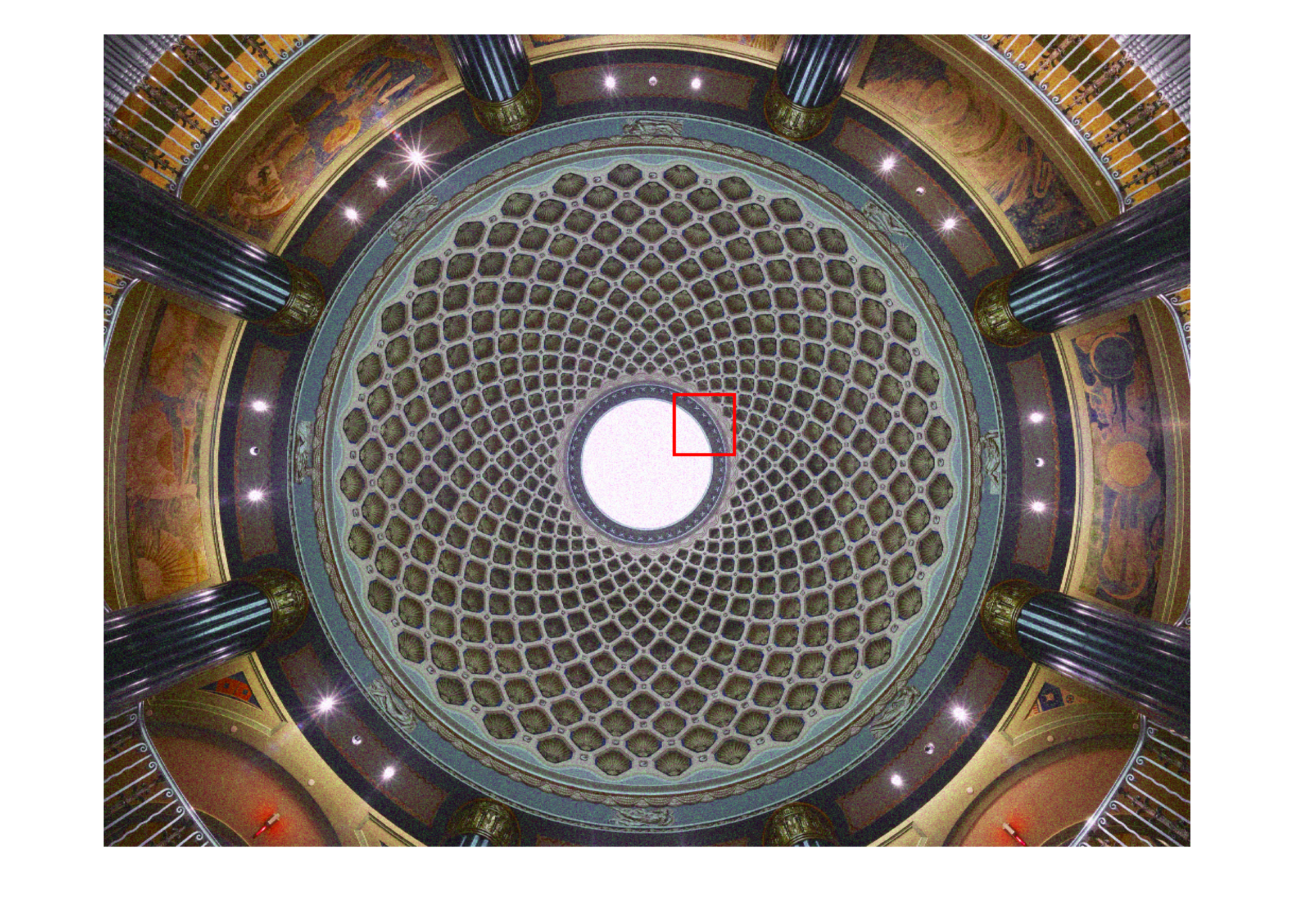}
    \end{minipage}
    \begin{minipage}[b]{0.45\linewidth}
    \centering
    \includegraphics[width=4.8cm]{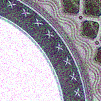}
   \vspace{11pt}
    \end{minipage}
    \hfill
    
    \centering
    \begin{minipage}[b]{0.45\linewidth}
    \centering
    \includegraphics[width=0.95\textwidth]{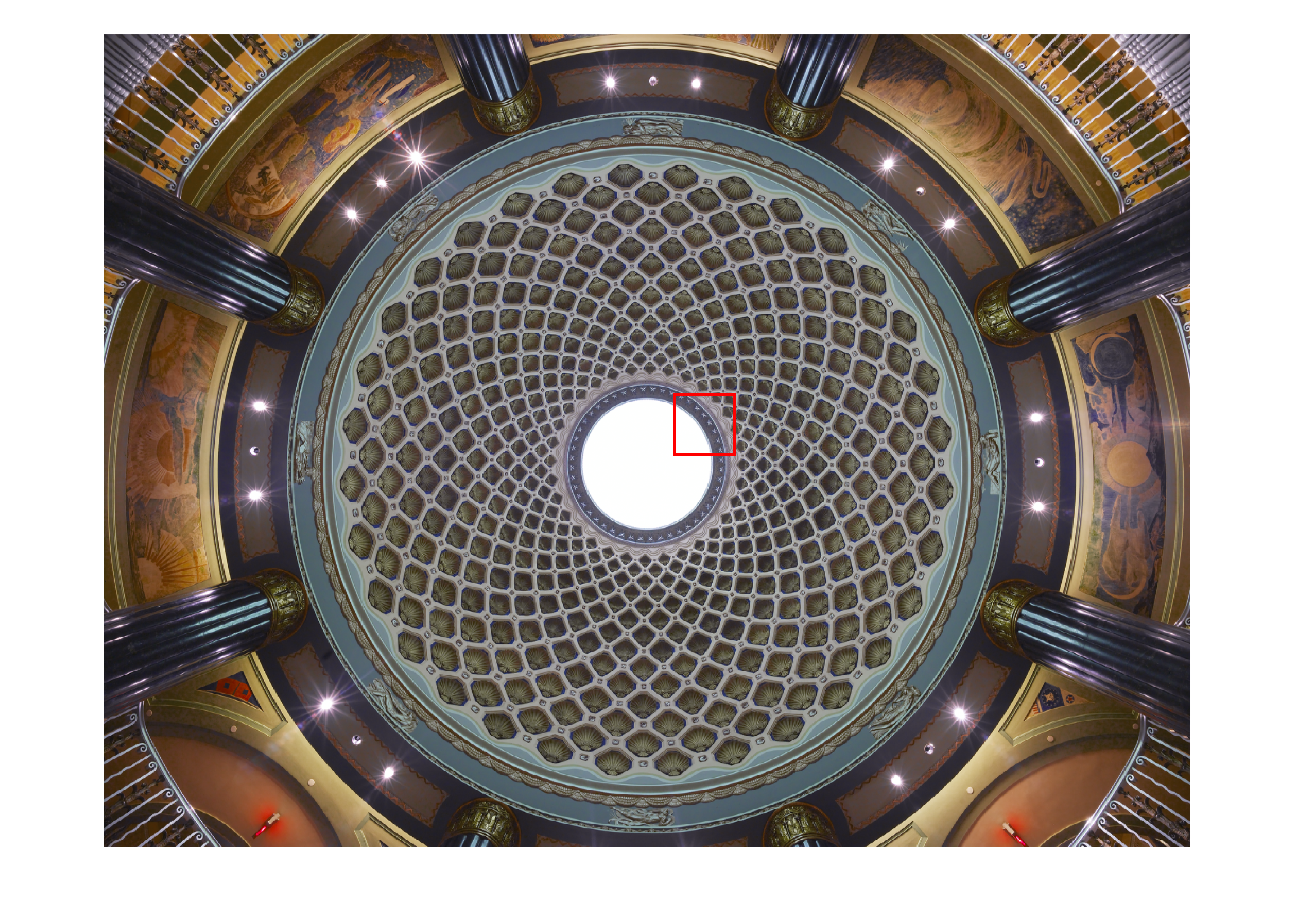}
    \end{minipage}
    \begin{minipage}[b]{0.45\linewidth}
    \centering
    \includegraphics[width=4.8cm]{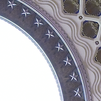}
   \vspace{11pt}
    \end{minipage}
    \hfill
    
    \centering
    \begin{minipage}[b]{0.45\linewidth}
    \centering
    \includegraphics[width=0.95\textwidth]{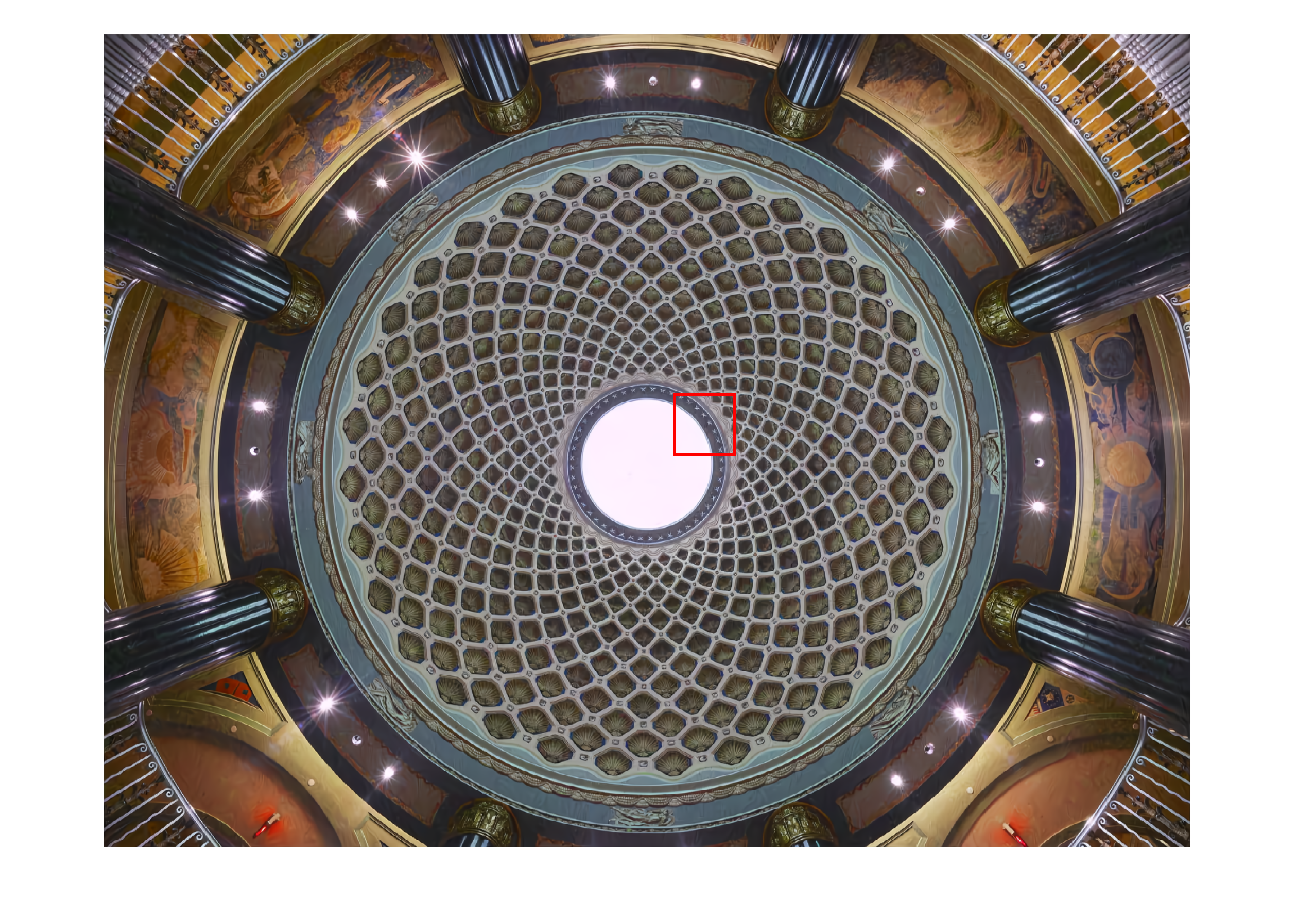}
    \end{minipage}
    \begin{minipage}[b]{0.45\linewidth}
    \centering
    \includegraphics[width=4.8cm]{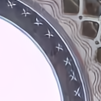}
    \vspace{11pt}
    \end{minipage}
    \hfill
    
    \centering
    \begin{minipage}[b]{0.45\linewidth}
    \centering
    \includegraphics[width=0.95\textwidth]{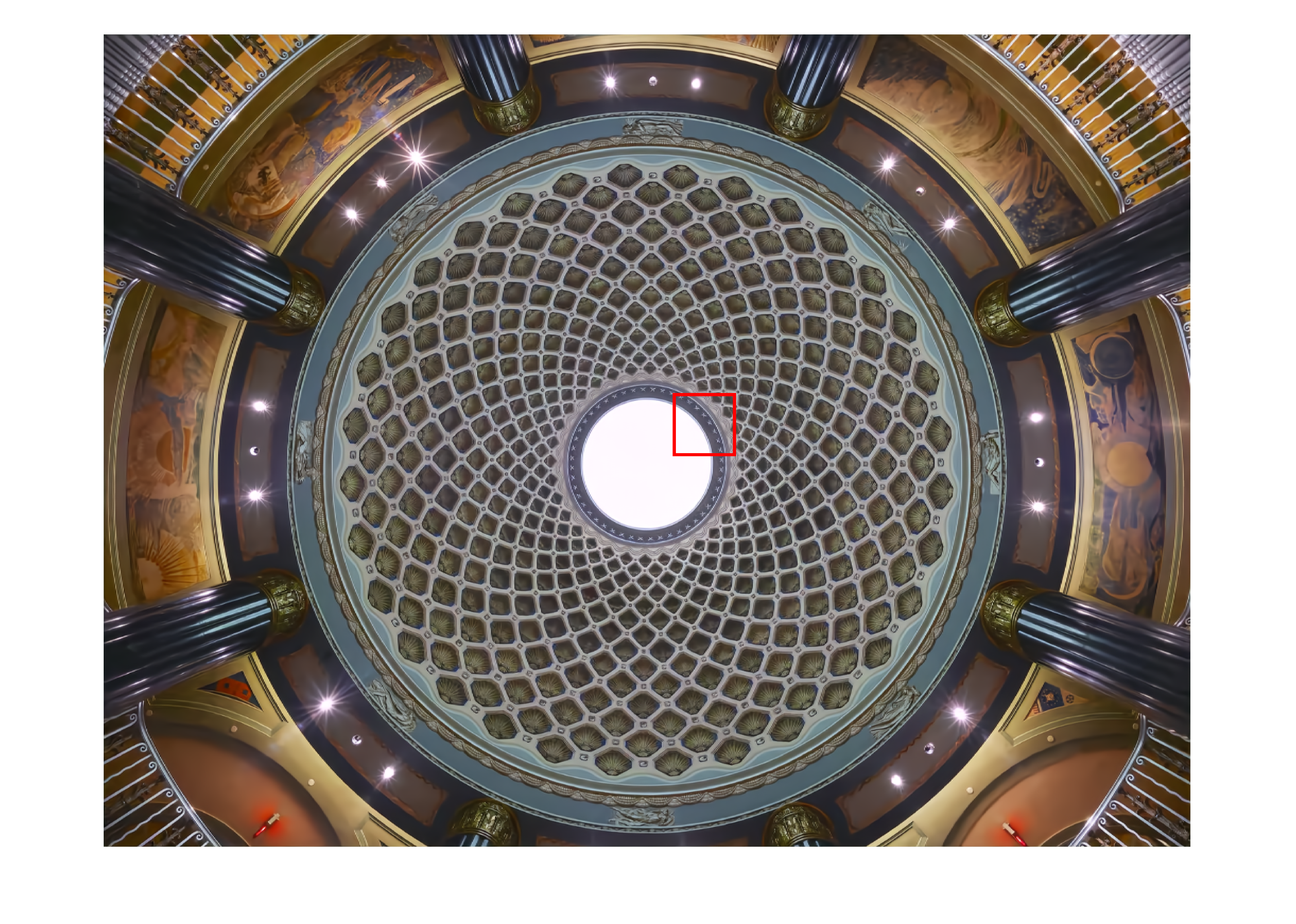}
    \end{minipage}
    \begin{minipage}[b]{0.45\linewidth}
    \centering
    \includegraphics[width=4.8cm]{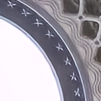}
    \vspace{11pt}
    \end{minipage}
\caption{An example of denoised images. Top to Bottom: noisy image, clean image, Denoised: Cheng+FFDNet-clip (bpp=0.57), Denoised: proposed (bpp=0.55). Images in the right column show the red square in the left images}
\label{fig:example_denoise}
\end{figure*}

A visual example comparing the denoised images obtained from JICD and Cheng+FFDNet-clip encoded at similar bitrates is shown in Fig.~\ref{fig:example_denoise}. As  seen in the figure, JICD preserves more details compared to Cheng+FFDNet-clip.  In addition, the colors inside the white circle are reproduced closer to the ground truth with JICD compared to the image produced by Cheng+FFDNet-clip.

\subsubsection{Noisy Image Reconstruction}
\label{subsubsec:exp_unseen_noisy_rec}
\begin{figure*}[t]
    \centering
    \begin{minipage}[b]{0.45\linewidth}
    \centering
    \includegraphics[width=\textwidth]{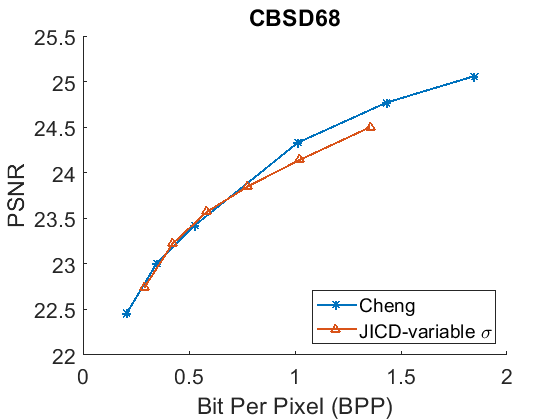}
    \centerline{(a)}\medskip
    \end{minipage}
    \begin{minipage}[b]{0.45\linewidth}
    \centering
    \includegraphics[width=\textwidth]{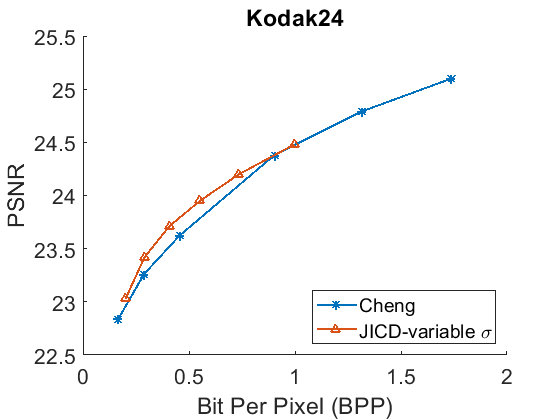}
    \centerline{(b)}\medskip
    \end{minipage}
    \hfill
    
    \centering
    \begin{minipage}[b]{0.45\linewidth}
    \centering
    \includegraphics[width=\textwidth]{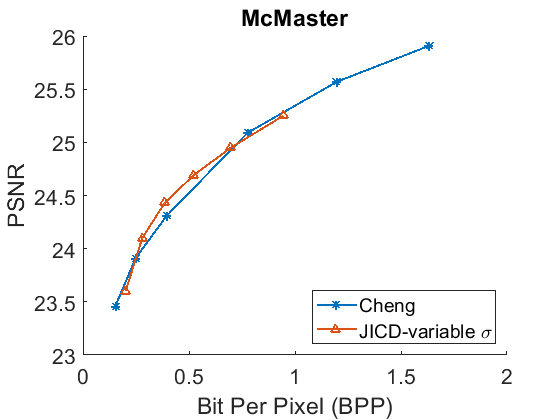}
    \centerline{(c)}\medskip
    \end{minipage}
    \begin{minipage}[b]{0.45\linewidth}
    \centering
    \includegraphics[width=\textwidth]{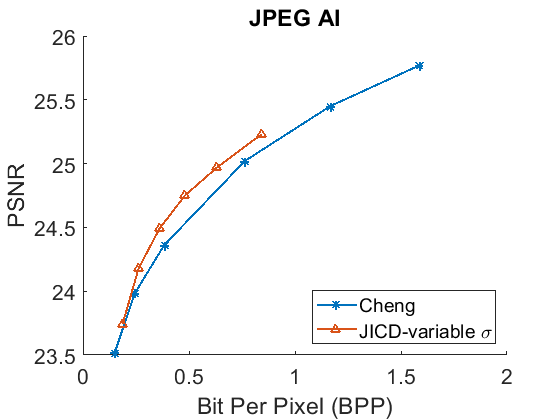}
    \centerline{(d)}\medskip
    \end{minipage}
    \hfill
\caption{The rate-PSNR curves for noisy input reconstruction. (a)~CBSD68, (b)~Kodak24, (c)~McMaster, (d)~JPEG-AI  }
\label{fig:results_noisy_png}
\end{figure*}

\begin{figure*}[t]
    \centering
    \begin{minipage}[b]{0.45\linewidth}
    \centering
    \includegraphics[width=\textwidth]{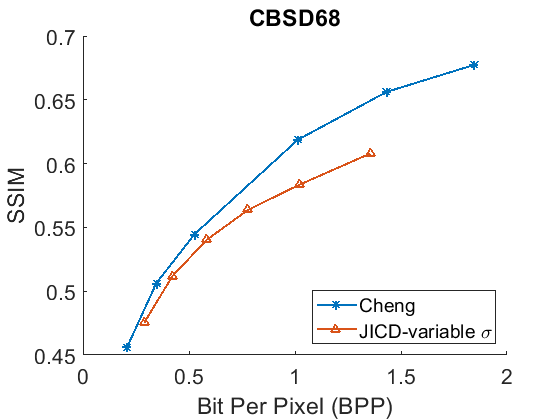}
    \centerline{(a)}\medskip
    \end{minipage}
    \begin{minipage}[b]{0.45\linewidth}
    \centering
    \includegraphics[width=\textwidth]{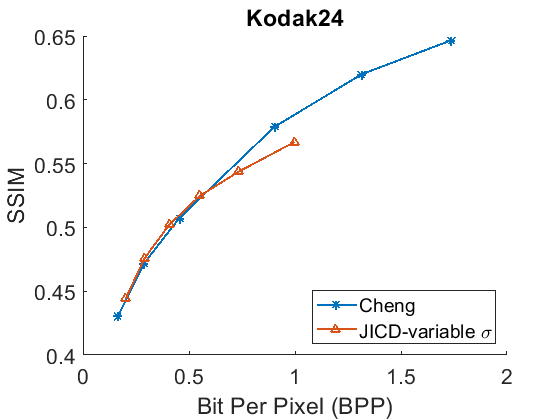}
    \centerline{(b)}\medskip
    \end{minipage}
    \hfill
    
    \centering
    \begin{minipage}[b]{0.45\linewidth}
    \centering
    \includegraphics[width=\textwidth]{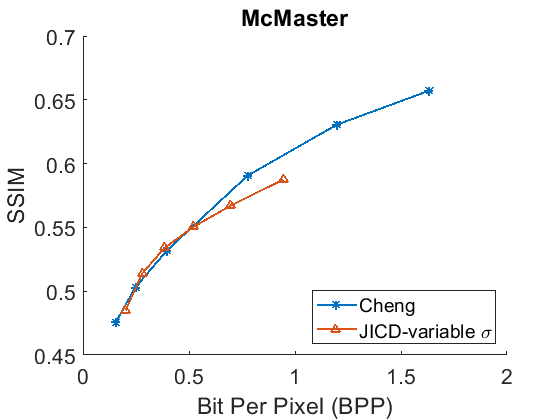}
    \centerline{(c)}\medskip
    \end{minipage}
    \begin{minipage}[b]{0.45\linewidth}
    \centering
    \includegraphics[width=\textwidth]{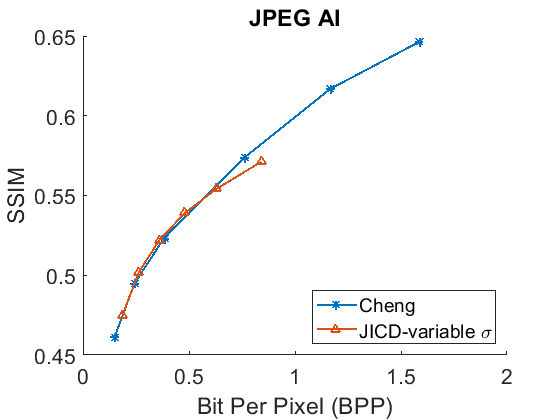}
    \centerline{(d)}\medskip
    \end{minipage}
    \hfill
\caption{{The rate-SSIM curves for noisy input reconstruction. (a)~CBSD68, (b)~Kodak24, (c)~McMaster, (d)~JPEG-AI}}
\label{fig:results_noisy_png_SSIM}
\end{figure*}

Besides denoising, the proposed JICD framework is also able to reconstruct the noisy input image when enhancement features are decoded together with base features. While the main focus of this work was on denoising (and the majority of experiments devoted to that goal), for completeness we also evaluate the noisy image reconstruction performance using unseen noise. We compare the noisy input reconstruction performance of JICD against~\cite{cheng2020image}, i.e., the compression model used earlier in the Cheng+FFDNet-clip baseline. The PSNR between the noisy input and the reconstructed noisy images is shown against bitrate in Fig.~\ref{fig:results_noisy_png}, {while Fig.~\ref{fig:results_noisy_png_SSIM} shows SSIM vs. bitrate}. As illustrated in Fig.~\ref{fig:results_noisy_png}, our JICD achieves better noisy input reconstruction compared to~\cite{cheng2020image} in most cases. BD-rate results corresponding to Fig.~\ref{fig:results_noisy_png} {and Fig.~\ref{fig:results_noisy_png_SSIM}} are given in Table~\ref{tbl:bd_rate_noisy} {and Table~\ref{tbl:bd_rate_noisy_SSIM}, respectively}. As the numbers in the Table~\ref{tbl:bd_rate_noisy} indicate, the proposed JICD achieves noticeable BD-rate savings on three of the four test datasets; the only exception is, again, the low-resolution CBSD68 dataset, where the loss is mainly concentrated at higher bitrates. { It is
worth noting that, since our proposed method is trained using the MSE loss, it performs better in
terms of PSNR than SSIM}. Overall, the proposed JICD framework achieves gains on both denoising and compression tasks compared to Cheng+FFDNet-clip and~\cite{cheng2020image} models. 


\renewcommand{\arraystretch}{1.2}
\begin{table*}[th]
\centering
\begin{tabular}{|c|crrrr|}
\hline
\multicolumn{1}{|c|}{Noise type}                                              &   Model                 & \multicolumn{1}{c}{CBSD68} & \multicolumn{1}{c}{Kodak24} & \multicolumn{1}{l}{McMaster} & JPEG AI \\ 
\hline \hline
\begin{tabular}[c]{@{}c@{}}Practical noise\\ simulator\end{tabular} & variable  $\sigma$ & $5.50$\%     &   $-11.74$\%     &   $-3.97$\%          &  $-13.49$\%
      \\ \hline
\end{tabular}
\caption{The PSNR-based BD-rate of the proposed JICD compared to the Cheng model on noisy input reconstruction.}
\label{tbl:bd_rate_noisy}
\end{table*}
\renewcommand{\arraystretch}{1}

\renewcommand{\arraystretch}{1.2}
\begin{table*}[th]
\centering
\begin{tabular}{|c|crrrr|}
\hline
\multicolumn{1}{|c|}{Noise type}                                              &   Model                 & \multicolumn{1}{c}{CBSD68} & \multicolumn{1}{c}{Kodak24} & \multicolumn{1}{l}{McMaster} & JPEG AI \\ 
\hline \hline
\begin{tabular}[c]{@{}c@{}}Practical noise\\ simulator\end{tabular} & variable  $\sigma$ & $22.58$\%     &   $1.90$\%     &   $4.05$\%          &  $0.58$\%
      \\ \hline
\end{tabular}
\caption{{The SSIM-based BD-rate of the proposed JICD compared to the Cheng model on noisy input reconstruction.}}
\label{tbl:bd_rate_noisy_SSIM}
\end{table*}
\renewcommand{\arraystretch}{1}

\section{Conclusion}
\label{sec:conclusion}
In this work, we presented a joint image compression and denoising framework. The proposed framework is a scalable multi-task image compression model based on the latent-space scalability. The base features are used to perform the denoising and the enhancement features are used when the noisy input reconstruction is needed. Extensive experiments show that the proposed framework achieves significant BD-rate savings up to 80.20\% across different dataset compared to the cascade compression and denoising method. The experimental results also indicate that the proposed method achieves improved results for the unseen noise for both denoising and noisy input reconstruction tasks.

\section*{Conflict of Interest Statement}

The authors declare that the research was conducted in the absence of any commercial or financial relationships that could be construed as a potential conflict of interest.

\section*{Author Contributions}

SRA and IVB contributed to conception and design of the study. HC developed the initial code. MU and SRA contributed to further code development and optimization. SRA wrote the first draft of the manuscript and worked with IVB on the revisions. 

\section*{Funding}
Funding for this work was provided by the Natural Sciences and Engineering Research Council  (NSERC) of Canada under the grants RGPIN-2021-02485 and RGPAS-2021-00038, and by Huawei Technologies.




\bibliographystyle{IEEEtran}
\bibliography{IEEEabrv,ref}



\end{document}